\documentclass[aps]{revtex4}  
\usepackage{graphicx}  
\usepackage{dcolumn}   
\usepackage{bm}        
\usepackage{amssymb}   
\usepackage{rotating}  
\def\MET{{\mbox{$E\kern-0.57em\raise0.19ex\hbox{/}_{T}$}}}
\def\met{{\mbox{$E\kern-0.57em\raise0.19ex\hbox{/}_{T}$}}}
\def\DZ{D0 }
\def\DZero{D0 }
\def\Dzero{D0 }

\def\ifb{~fb$^{-1}$}
\def\pp{$p\bar{p}$}

\def\ttbar{$t\bar{t}$}
\def\WH{$WH\rightarrow \ell\nu b\bar{b}$}

\def\lmet{$WH\rightarrow \ell\kern-0.45em\raise0.19ex\hbox{/} \nu b\bar{b}$}
\def\ZH{$ZH\rightarrow \nu\bar{\nu} b\bar{b}$}
\def\ZHll{$ZH\rightarrow \ell^+ \ell^- b\bar{b}$}

\def\www{$WH \rightarrow WW^{+} W^{-}$}

\def\hww{$H\rightarrow W^+ W^-$}
\def\hbb{$H\rightarrow b\bar{b}$}

\def\tevE{$\sqrt{s}=1.96$~TeV}

\begin{document}
\rightline{FERMILAB-CONF-09-557-E}
\rightline{CDF Note 9998}
\rightline{\DZ Note 5983}
\vskip0.5in

\title{Combined CDF and \DZ Upper Limits on Standard Model Higgs-Boson Production with 2.1 - 5.4 fb$^{-1}$ of Data\\[2.5cm]}

\author{
The TEVNPH Working Group\footnote{The Tevatron
New-Phenomena and Higgs working group can be contacted at
TEVNPHWG@fnal.gov. More information can be found at http://tevnphwg.fnal.gov/.}
 } 
\affiliation{\vskip0.3cm for the CDF and \DZ Collaborations\\ \vskip0.2cm
\today} 
\begin{abstract} 
\vskip0.3in 
We combine results from CDF and D0 on direct searches for a standard
model (SM) Higgs boson ($H$) in \pp~collisions at the Fermilab
Tevatron at $\sqrt{s}=1.96$~TeV. Compared to the previous Tevatron Higgs
search combination more data 
have been added and some previously used channels have been
reanalyzed to gain sensitivity. We use the latest parton distribution
functions and $gg \rightarrow H$ theoretical cross sections when
comparing our limits to the SM predictions. 
With 2.0-4.8\ifb\ of data analyzed at CDF, and 2.1-5.4\ifb\ at D0, 
the 95\% C.L. upper limits on Higgs boson production are a factor of 
2.70~(0.94) times the SM
cross section for a Higgs boson mass of $m_{H}=$115~(165)~GeV/$c^2$.
The corresponding median upper limits expected in the absence of Higgs boson production
are 1.78~(0.89).  The mass range excluded at 95\% C.L. for a SM Higgs 
is  $163<m_{H}<166$~GeV/$c^{2}$, with an expected exclusion of $159<m_{H}<168$~GeV/$c^{2}$.
\\[2cm]
{\hspace*{5.5cm}\em Preliminary Results}
\end{abstract}

\maketitle

\newpage
\section{Introduction} 

The search for a mechanism for electroweak symmetry breaking, and in
particular for a standard model (SM) Higgs boson has been a major goal
of particle physics for many years, and is a central part of the
Fermilab Tevatron physics program. Both the CDF and \Dzero experiments
have performed new combinations~\cite{CDFHiggs,DZHiggs} of multiple
direct searches for the SM Higgs boson.  The new searches include more
data and improved analysis techniques compared to previous analyses.
The sensitivities of these new combinations significantly exceed those of
previous combinations~\cite{prevhiggs}.

In this note, we combine the most recent results of all such 
searches in \pp~collisions at~\tevE.  The analyses combined 
here seek signals of Higgs bosons produced in association with 
vector bosons ($q\bar{q}\rightarrow W/ZH$), through gluon-gluon 
fusion ($gg\rightarrow H$), and through vector boson fusion (VBF) 
($q\bar{q}\rightarrow q^{\prime}\bar{q}^{\prime}H$) corresponding 
to integrated luminosities ranging from 2.0 to 4.8\ifb~at CDF and 
2.1 to 5.4\ifb~at D0.  The Higgs boson decay modes studied are
$H\rightarrow b{\bar{b}}$, $H\rightarrow W^+W^-$, $H\rightarrow
\tau^+\tau^-$ and $H\rightarrow \gamma\gamma$.

To simplify the combination, the searches are separated into 90
mutually exclusive final states (36 for CDF and 54 for D0; see
Table~\ref{tab:cdfacc} and ~\ref{tab:dzacc}) referred to as
``analyses'' in this note.  The selection procedures for each 
analysis are detailed in Refs.~\cite{cdfWH2J} through~\cite{dzttH}, 
and are briefly described below.

\section{Acceptance, Backgrounds, and Luminosity}  

Event selections are similar for the corresponding CDF and D0 analyses.
For the case of \WH, an isolated lepton ($\ell=$ electron or muon) and 
two jets are required, with one or more $b$-tagged jet, i.e., identified 
as containing a weakly-decaying $B$ hadron.  
Selected events must also display a significant imbalance 
in transverse momentum
(referred to as missing transverse energy or \met).  Events with more than one
isolated lepton are vetoed.  
For the D0 \WH\ analyses, two and three jet events are analyzed separately, and 
in each of these samples
two non-overlapping $b$-tagged samples are
defined, one being a single ``tight'' $b$-tag (ST) sample, and the other a 
double ``loose'' $b$-tag (DT) sample. The tight and loose $b$-tagging criteria
are defined with respect to the mis-identification 
rate that the $b$-tagging algorithm yields for light quark or gluon jets 
(``mistag rate'') typically $\le 
0.5\%$ or $\le 1.5\%$, respectively.  
The final variable is a neural network output which takes as input seven kinematic
variables for the two-jet sample, while for the
three-jet sample the dijet invariant mass is used.
%
%

For the CDF \WH\ analyses, the events are analyzed in two- and three-jet subsamples
 separately, and in each of these samples the events are grouped
into various lepton and $b$-tag categories. In addition to the
selections requiring an identified lepton, events with an 
isolated track failing lepton selection requirements in the two-jet sample, or an
identified loose muon in the extended muon coverage in the three-jet sample, are
grouped into their own categories. This provides some acceptance for
single prong tau decays. Within the lepton categories there 
are four $b$-tagging categories considered in the two-jets sample -- two
tight $b$-tags (TDT), one tight $b$-tag and one loose $b$-tag (LDT), one
tight $b$-tag and one looser $b$-tag (LDTX), and a single, tight, $b$-
tag (ST). These $b$-tag category names are also used in the three-jets,
$\MET b{\bar{b}}$, and $\ell^+\ell^-b{\bar{b}}$ channel descriptions,
except the LDTX events are included in ST events where the looser $b$-
tag is ignored.  A Bayesian neural network discriminant is trained at
each $m_H$ in the test range for the two-jet sample, separately for each
category, while for the three-jet sample a matrix element discriminant
is used.

For the \ZH\ analyses, the selection is
similar to the $WH$ selection, except all events with isolated leptons are vetoed and
stronger multijet background suppression techniques are applied. 
Both CDF and D0  analyses use a track-based missing transverse momentum calculation
as a discriminant against false \met . In addition, D0 trains a boosted decision tree to discriminate against 
the multijet background.  There is a sizable fraction of the \WH\ signal in which the lepton is undetected
that is selected in the \ZH\ samples,  so these analyses are also referred to as
$VH \rightarrow \met b \bar{b}$.
The CDF  analysis uses three non-overlapping samples of 
events (TDT, LDT and ST as for $WH$). 
D0 uses orthogonal ST and tight-loose double-tag (TLDT) channels.
CDF uses
neural-network discriminants as the final variables, while D0 
uses boosted decision trees as the advanced analysis technique.

The \ZHll\ analyses require two isolated leptons and
at least two jets.   D0's  \ZHll\ analyses separate events into non-overlapping samples of
events with one tight $b$-tag (ST) and two loose $b$-tags (DT).  CDF separates events into
ST, TDT, and LDT samples. 
For the D0 analysis boosted decision trees provide the 
final variables for setting  limits, while CDF uses the output
of a two-dimensional neural network. For this combination D0 has increased the
signal acceptance by loosening the selection criteria for one of the leptons. In addition a kinematic
fit is now applied to the $Z-$boson and jets.   
CDF corrects jet energies for \met\ using a neural network approach.
In this analysis the events are
divided  into three tagging
categories: tight double tags, loose double tags, and single tags.  Both CDF and D0 further
subdivide the channels into lepton categories with different signal-to-background characteristics.

For the \hww~analyses, signal events are characterized by
a large \met~and two opposite-signed, isolated leptons.  The presence of
neutrinos in the final state prevents the reconstruction of the
candidate Higgs boson mass. 
 D0 selects events containing electrons and muons,
dividing the data sample into three final states:
$e^+e^-$, $e^\pm \mu^\mp$, and $\mu^+\mu^-$.
CDF separates the \hww\ events in six non-overlapping samples,
labeled ``high $s/b$'' and ``low $s/b$'' for the lepton selection categories, and
also split by the number of jets: 0, 1, or 2+ jets.  The sample with two or more jets
is not split into low $s/b$ and high $s/b$ lepton categories.   
The sixth CDF channel is a new low dilepton mass ($m_{\ell^+\ell^-}$) channel,
which accepts events with $m_{\ell^+\ell^-}<16$~GeV.  This channel increases the sensitivity of the
$H\rightarrow W^+W^-$ analyses at low $m_H$, adding 10\% additional acceptance at $m_H=120$~GeV.
CDF's division of events
into jet categories allows the analysis discriminants to separate three different categories of
signals from the backgrounds more effectively.  
The signal production mechanisms considered are
$gg\rightarrow H\rightarrow W^+W^-$, $WH+ZH\rightarrow jjW^+W^-$, and the vector-boson
fusion process.  For $gg\rightarrow H$, however, recent work~\cite{anastasiouwebber} indicates that
the theoretical uncertainties due to scale and PDF variations are significantly different in
the different jet categories.  CDF and D0 now divide the theoretical uncertainty on $gg\rightarrow H$
into PDF and scale pieces, and use the differential uncertainties of~\cite{anastasiouwebber}.
D0 uses neural-networks, including the number of jets as an input, as the final discriminant.
CDF likewise uses neural-networks, including likelihoods constructed from matrix-element probabilities (ME) as input
in the 0-jet bin.
All analyses in this channel have been updated with more data and analysis improvements.

The CDF collaboration also contributes an analysis searching for Higgs bosons decaying
to a tau lepton pair, in three separate production channels: 
direct $gg \rightarrow H$ production, associated $WH$ or $ZH$ production,
or vector boson production with $H$ and forward jets in the final state.  Two jets are required
in the event selection.  
In  this analysis, the final variable for setting  limits is
a combination of several neural-network discriminants.   The theoretical systematic uncertainty
on the $gg \rightarrow H$ production rate now takes into account recent theoretical work~\cite{anastasiouwebber}
which provides uncertainties in each jet category.

D0 also
contributes an analysis 
for the final state $\tau
\tau$ jet jet, which is sensitive to the $VH\rightarrow jj \tau \tau$, $ZH 
\rightarrow \tau \tau b \bar{b}$, VBF and gluon gluon fusion
(with two additional jets) mechanisms.  A neural network output is used as the discriminant variable for RunIIa (the first 
1.0 fb$^{-1}$ of data), 
while a boosted decision tree output is used for later data. 

The CDF collaboration introduces a new all-hadronic channel, $WH+ZH\rightarrow jjb{\bar{b}}$ for this
combination.  Events are selected with four jets, at least two of which are $b$-tagged
with the tight $b$-tagger.  The large QCD backgrounds are estimated with the use of
data control samples, and the final variable is a matrix element signal probability
discriminant.

The D0 collaboration
contributes three \www\ analyses, where the associated $W$ boson and
the $W$ boson from the Higgs boson decay that has the same charge are required 
to decay leptonically,
thereby defining three like-sign dilepton final states 
($e^\pm e^\pm$, $e^\pm \mu^\pm$, and $\mu^{\pm}\mu^{\pm}$)
containing all decays of the third
$W$ boson. In  this analysis the final variable is a likelihood discriminant formed from several
topological variables.   
CDF contributes a \www\ analysis using a selection of like-sign
dileptons and a neural network to further purify the signal.

D0 also contributes an analysis searching for direct Higgs boson production
decaying to a photon pair in 4.2 fb$^{-1}$ of data.
In  this analysis, the final variable is the invariant mass of the two-photon system.
Finally, D0 includes the channel
$t \bar{t} H \rightarrow t \bar{t} b \bar{b}$.  Here the samples are analyzed 
independently according to the
number of $b$-tagged jets (1,2,3, i.e. ST,DT,TT) 
and the total number of jets (4 or 5).
The total transverse energy of the reconstructed objects ($H_T$) is used
as discriminant variable.

We normalize our Higgs boson signal predictions to the most recent high-order calculations
available.
The $gg\rightarrow H$ production cross section is calculated at NNLL in
QCD and also includes two-loop electroweak
effects; see Refs.~\cite{anastasiou,grazzinideflorian} and references therein for 
the different steps of these calculations.   The newer calculation
includes a more thorough treatment of higher-order radiative corrections, particularly those involving
$b$ quark loops.  The $gg\rightarrow H$ production cross section depends strongly on the PDF set chosen and the
accompanying value of $\alpha_s$.  The cross sections used 
here are calculated with the MSTW 2008 NNLO PDF set~\cite{mstw2008}.   The new $gg\rightarrow H$ cross sections
supersede those used in the update of Summer 2008~\cite{TevHiggsICHEP,nnlo1,aglietti}, which had a simpler
treatment of radiative corrections and used the older MRST 2002 PDF set~\cite{mrst2002}.  The Higgs boson
production cross sections used here
are listed in Table~\ref{tab:higgsxsec}~\cite{grazzinideflorian}.  Furthermore, we now include the larger
theoretical uncertainties due to scale variations and PDF variations separately for each jet bin for
the $gg\rightarrow H$ processes as evaluated in~\cite{anastasiouwebber}.  We treat the scale uncertainties
as 100\% correlated between jet bins and between CDF and D0, and also treat the PDF uncertainties in the
cross section as correlated between jet bins and between CDF and D0.  
We include all significant Higgs production modes in the high mass search.   Besides gluon-gluon 
fusion through a virtual top quark loop  (ggH), 
we include production in association 
with a $W$ or $Z$ vector boson  (VH)~\cite{nnlo2,Brein,Ciccolini}, 
and vector boson  fusion (VBF)~\cite{nnlo2,Berger}.
In order to predict the distributions of the kinematics of Higgs boson signal events, CDF and D0 
 use the \textsc{PYTHIA}~\cite{pythia} Monte Carlo program, with
\textsc{CTEQ5L} or \textsc{CTEQ6L}~\cite{cteq} leading-order (LO)
parton distribution functions.  
The Higgs boson decay branching ratio predictions are calculated with HDECAY~\cite{hdecay}, and are
also listed in Table~\ref{tab:higgsxsec}.

For both CDF and D0, events from
multijet (instrumental) backgrounds (``QCD production'') are measured
in independent data samples using several different methods.
For CDF, backgrounds
from SM processes with electroweak gauge bosons or top quarks were generated using \textsc{PYTHIA},
\textsc{ALPGEN}~\cite{alpgen}, \textsc{MC@NLO}~\cite{MC@NLO}
 and \textsc{HERWIG}~\cite{herwig}
programs. For D0, these backgrounds were generated using
\textsc{PYTHIA}, \textsc{ALPGEN}, and \textsc{COMPHEP}~\cite{comphep},
with \textsc{PYTHIA} providing parton-showering and hadronization for
all the generators.  These background processes were normalized using either
experimental data or next-to-leading order calculations (including
\textsc{MCFM}~\cite{mcfm} for $W+$ heavy flavor process).

\begin{table}
\caption{
The (N)NLO production cross sections and decay branching fractions for the SM
Higgs boson assumed for the combination}
\vspace{0.2cm}
\label{tab:higgsxsec}
\begin{ruledtabular}
\begin{tabular}{cccccccc}\\
$m_H$ & $\sigma_{gg\rightarrow H}$ & $\sigma_{WH}$ & $\sigma_{ZH}$ & $\sigma_{\rm{VBF}}$ &  
$B(H\rightarrow b{\bar{b}})$ & $B(H\rightarrow \tau^+{\tau^-})$ & $B(H\rightarrow W^+W^-)$ \\ 
(GeV/$c^2$) & (fb) & (fb) & (fb) & (fb) & (\%) & (\%) & (\%) \\ \hline
   100 &    1861   &   286.1  &   166.7  &   99.5  &  81.21 & 7.924 &  1.009 \\ 
   105 &    1618   &   244.6  &   144.0  &   93.3  &  79.57 & 7.838 &  2.216 \\
   110 &    1413   &   209.2  &   124.3  &   87.1  &  77.02 & 7.656 &  4.411  \\
   115 &    1240   &   178.8  &   107.4  &   79.1  &  73.22 & 7.340 &  7.974 \\
   120 &    1093   &   152.9  &   92.7   &   71.6  &  67.89 & 6.861 &  13.20 \\
   125 &    967    &   132.4  &   81.1   &   67.4  &  60.97 & 6.210 &  20.18 \\
   130 &    858    &   114.7  &   70.9   &   62.5  &  52.71 & 5.408 &  28.69 \\
   135 &    764    &   99.3   &   62.0   &   57.6  &  43.62 & 4.507 &  38.28 \\
   140 &    682    &   86.0   &   54.2   &   52.6  &  34.36 & 3.574 &  48.33 \\
   145 &    611    &   75.3   &   48.0   &   49.2  &  25.56 & 2.676 &  58.33 \\
   150 &    548    &   66.0   &   42.5   &   45.7  &  17.57 & 1.851 &  68.17 \\
   155 &    492    &   57.8   &   37.6   &   42.2  &  10.49 & 1.112 &  78.23 \\
   160 &    439    &   50.7   &   33.3   &   38.6  &  4.00  & 0.426 &  90.11 \\
   165 &    389    &   44.4   &   29.5   &   36.1  &  1.265 & 0.136 &  96.10 \\
   170 &    349    &   38.9   &   26.1   &   33.6  &  0.846 & 0.091 &  96.53 \\
   175 &    314    &   34.6   &   23.3   &   31.1  &  0.663 & 0.072 &  95.94 \\
   180 &    283    &   30.7   &   20.8   &   28.6  &  0.541 & 0.059 &  93.45 \\
   185 &    255    &   27.3   &   18.6   &   26.8  &  0.420 & 0.046 &  83.79 \\
   190 &    231    &   24.3   &   16.6   &   24.9  &  0.342 & 0.038 &  77.61 \\
   195 &    210    &   21.7   &   15.0   &   23.0  &  0.295 & 0.033 &  74.95 \\
   200 &    192    &   19.3   &   13.5   &   21.2 &  0.260 & 0.029 &  73.47 \\
\end{tabular}
\end{ruledtabular}
\end{table}

Tables~\ref{tab:cdfacc} and~\ref{tab:dzacc} summarize, for CDF and D0 respectively,
the integrated luminosities, the Higgs boson mass ranges over which the searches are performed,
and references to further details for each analysis.


\begin{table}[h]
\caption{\label{tab:cdfacc}Luminosity, explored mass range and references
for the different processes
and final state ($\ell=e, \mu$) for the CDF analyses
}
\begin{ruledtabular}
\begin{tabular}{lccc} \\
Channel & Luminosity (fb$^{-1}$) & $m_H$ range (GeV/$c^2$) & Reference \\ \hline
$WH\rightarrow \ell\nu b\bar{b}$ 2-jet channels\ \ \ 3$\times$(TDT,LDT,ST,LDTX)        & 4.3  & 100-150 & \cite{cdfWH2J} \\
$WH\rightarrow \ell\nu b\bar{b}$ 3-jet channels\ \ \ 2$\times$(TDT,LDT,ST)             & 4.3  & 100-150 & \cite{cdfWH3J} \\
$ZH\rightarrow \nu\bar{\nu} b\bar{b}$ \ \ \ (TDT,LDT,ST)            & 3.6  & 105-150 & \cite{cdfmetbb} \\
$ZH\rightarrow \ell^+\ell^- b\bar{b}$ \ \ \ (low,high $s/b$)$\times$(TDT,LDT,ST)   & 4.1  & 100-150 & \cite{cdfZHll} \\
$H\rightarrow W^+ W^-$ \ \ \ (low,high $s/b$)$\times$(0,1 jets)+(2+ jets)+Low-$m_{\ell\ell}$  & 4.8  & 110-200 & \cite{cdfHWW} \\
$WH \rightarrow WW^+ W^- \rightarrow \ell^\pm\nu \ell^\pm\nu$ & 4.8  & 110-200 & \cite{cdfHWW} \\
$H$ + $X\rightarrow \tau^+ \tau^- $ + 2 jets                  & 2.0  & 110-150 & \cite{cdfHtt} \\
$WH+ZH\rightarrow jjb{\bar{b}}$                               & 2.0  & 100-150 & \cite{cdfjjbb} \\ 
\end{tabular}
\end{ruledtabular}
\end{table}

\vglue 0.5cm 

\begin{table}[h]
\caption{\label{tab:dzacc}Luminosity, explored mass range and references 
for the different processes
and final state ($\ell=e, \mu$) for the D0 analyses
}
\begin{ruledtabular}
\begin{tabular}{lccc} \\
Channel & Luminosity (fb$^{-1}$) & $m_H$ range (GeV/$c^2$) & Reference \\ \hline
$WH\rightarrow \ell\nu b\bar{b}$ \ \ \ 2$\times$(ST,DT)             & 5.0  & 100-150 & \cite{dzWHl} \\
%
$VH\rightarrow \tau\tau b\bar{b}/q\bar{q} \tau\tau$           & 4.9  & 105-145 & \cite{dzVHt1,dzVHt2} \\
$ZH\rightarrow \nu\bar{\nu} b\bar{b}$ \ \ \ (ST,TLDT)                    & 5.2  & 100-150 & \cite{dzZHv} \\
$ZH\rightarrow \ell^+\ell^- b\bar{b}$ \ \ \ 2$\times$(ST,DT)        & 4.2  & 100-150 & \cite{dzZHll} \\
$WH \rightarrow WW^+ W^- \rightarrow \ell^\pm\nu \ell^\pm\nu$ & 3.6  & 120-200 & \cite{dzWWW1,dzWWW2} \\
$H\rightarrow W^+ W^- \rightarrow \ell^\pm\nu \ell^\mp\nu$    & 5.4  & 115-200 & \cite{dzHWW}\\
$H \rightarrow \gamma \gamma$                                 & 4.2  & 100-150 & \cite{dzHgg} \\ 
$t \bar{t} H \rightarrow t \bar{t} b \bar{b}$ \ \ \ 2$\times$(ST,DT,TT)& 2.1  & 105-155 & \cite{dzttH} \\ 
\end{tabular}
\end{ruledtabular}
\end{table}

\section{Distributions of Candidates} 

All analyses provide binned histograms of the final discriminant variables
for the signal and background predictions, itemized separately for each
source, and the data.
The number of channels combined is large, and the number of bins
in each channel is large.  Therefore, the task of assembling
histograms and checking whether the expected and observed limits are
consistent with the input predictions and observed data is difficult.
We therefore provide histograms that aggregate all channels' signal,
background, and data together.  In order to preserve most of the
sensitivity gain that is achieved by the analyses by binning the data
instead of collecting them all together and counting, we aggregate the
data and predictions in narrow bins of signal-to-background ratio,
$s/b$.  Data with similar $s/b$ may be added together with no loss in
sensitivity, assuming similar systematic errors on the predictions.
The aggregate histograms do not show the effects of systematic
uncertainties, but instead compare the data with the central
predictions supplied by each analysis.

The range of $s/b$ is quite large in each analysis, and so
$\log_{10}(s/b)$ is chosen as the plotting variable.  Plots of the
distributions of $\log_{10}(s/b)$ are shown for $m_H=115$ and
165~GeV/$c^2$ in Figure~\ref{fig:lnsb}.  These distributions can be
integrated from the high-$s/b$ side downwards, showing the sums of
signal, background, and data for the most pure portions of the
selection of all channels added together.  These integrals can be seen
in Figure~\ref{fig:integ}.  The most significant candidates are be found
in the bins with the highest $s/b$; an excess in these bins relative to the
background prediction drives the Higgs boson cross section limit upwards, while
a deficit drives it downwards.  The lower-$s/b$ bins show that the modeling of the
rates and kinematic distributions of the backgrounds is very good.
  The integrated plots
show the excess of events in the high-$s/b$ bins for the analyses seeking
a Higgs boson mass of 115~GeV/$c^2$, and a deficit of events in the high-$s/b$
bins for the analyses seeking a Higgs boson of mass 165~GeV/$c^2$.


 \begin{figure}[t]
 \begin{centering}
 \includegraphics[width=0.4\textwidth]{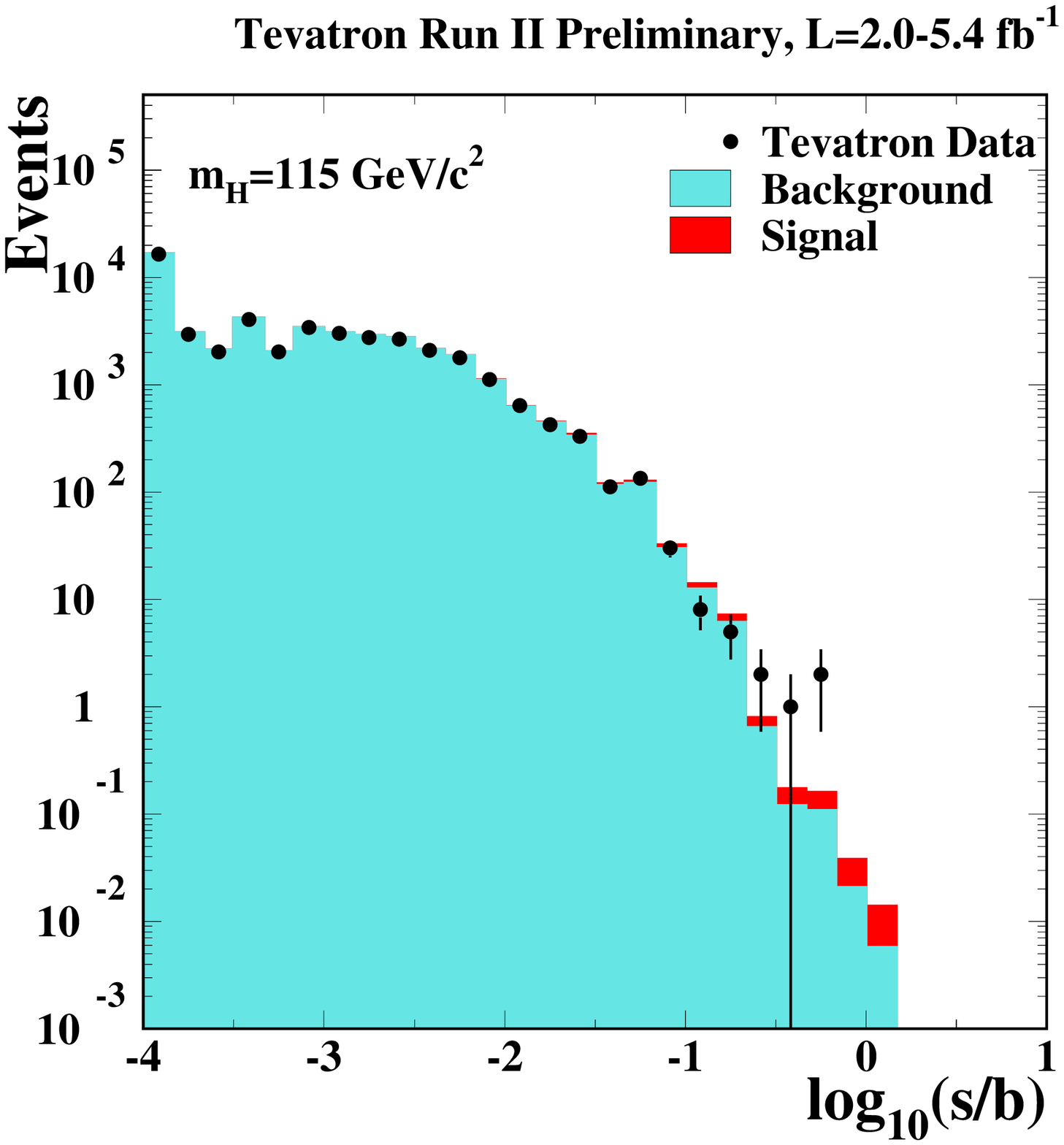}\includegraphics[width=0.4\textwidth]{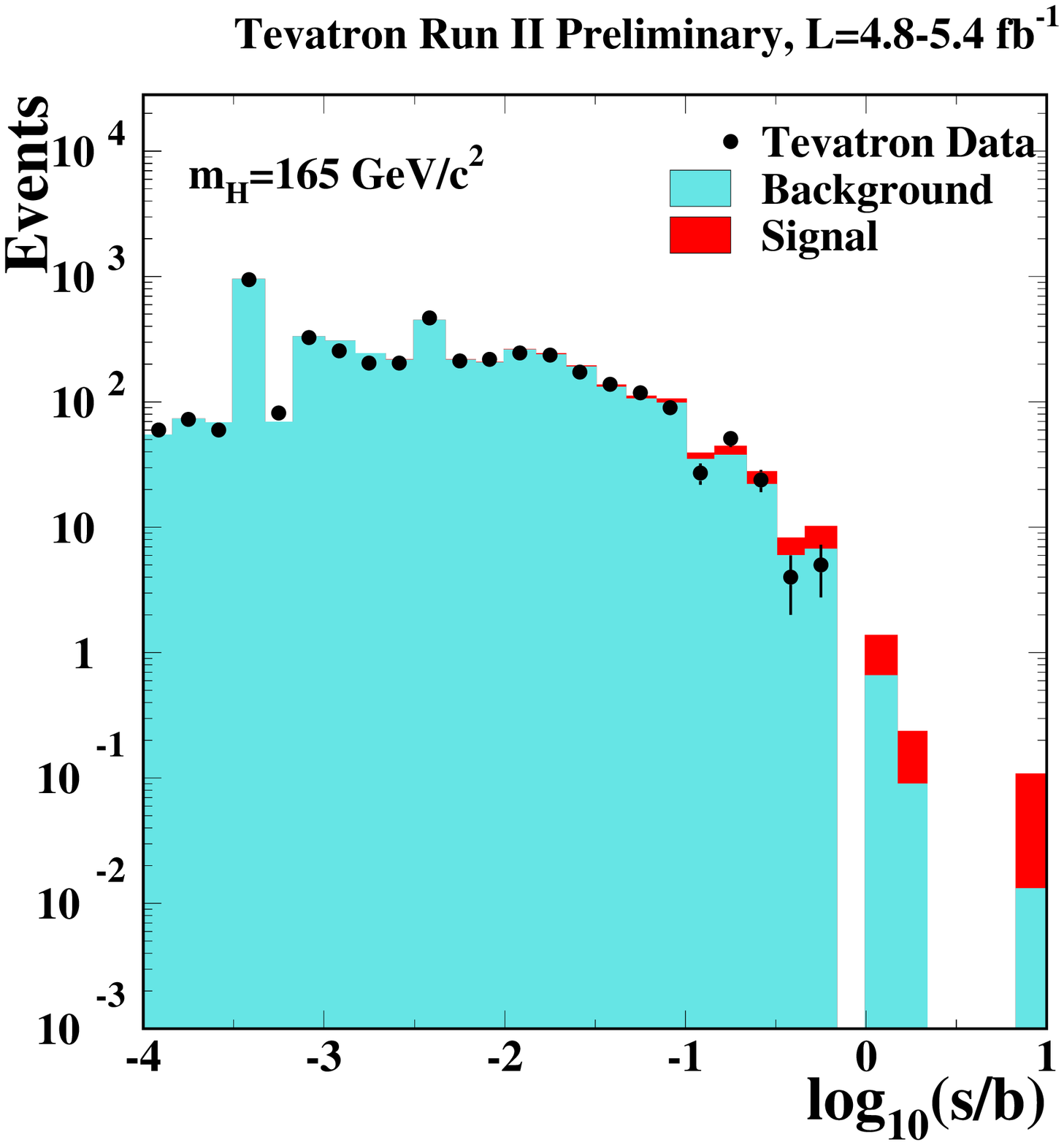}
 \caption{
 \label{fig:lnsb} Distributions of $\log_{10}(s/b)$, for the data from all contributing channels from
CDF and D0, for Higgs boson masses of 115 and 165~GeV/$c^2$.  The
data are shown with points, and the expected signal is shown stacked on top of
the backgrounds.  Underflows and overflows are collected into the
bottom and top bins. }
 \end{centering}
 \end{figure}

 \begin{figure}[t]
 \begin{centering}
 \includegraphics[width=0.4\textwidth]{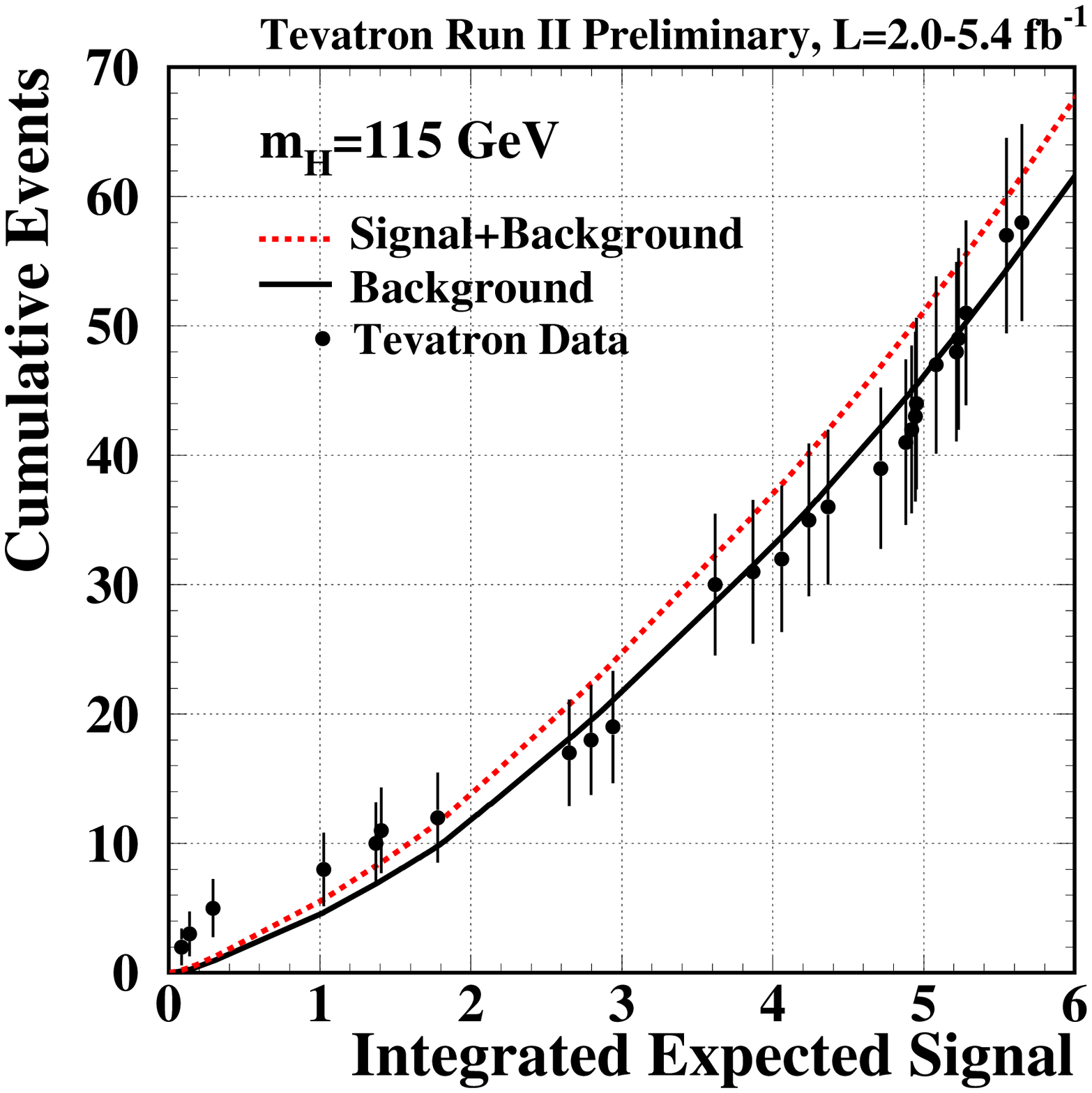}\includegraphics[width=0.4\textwidth]{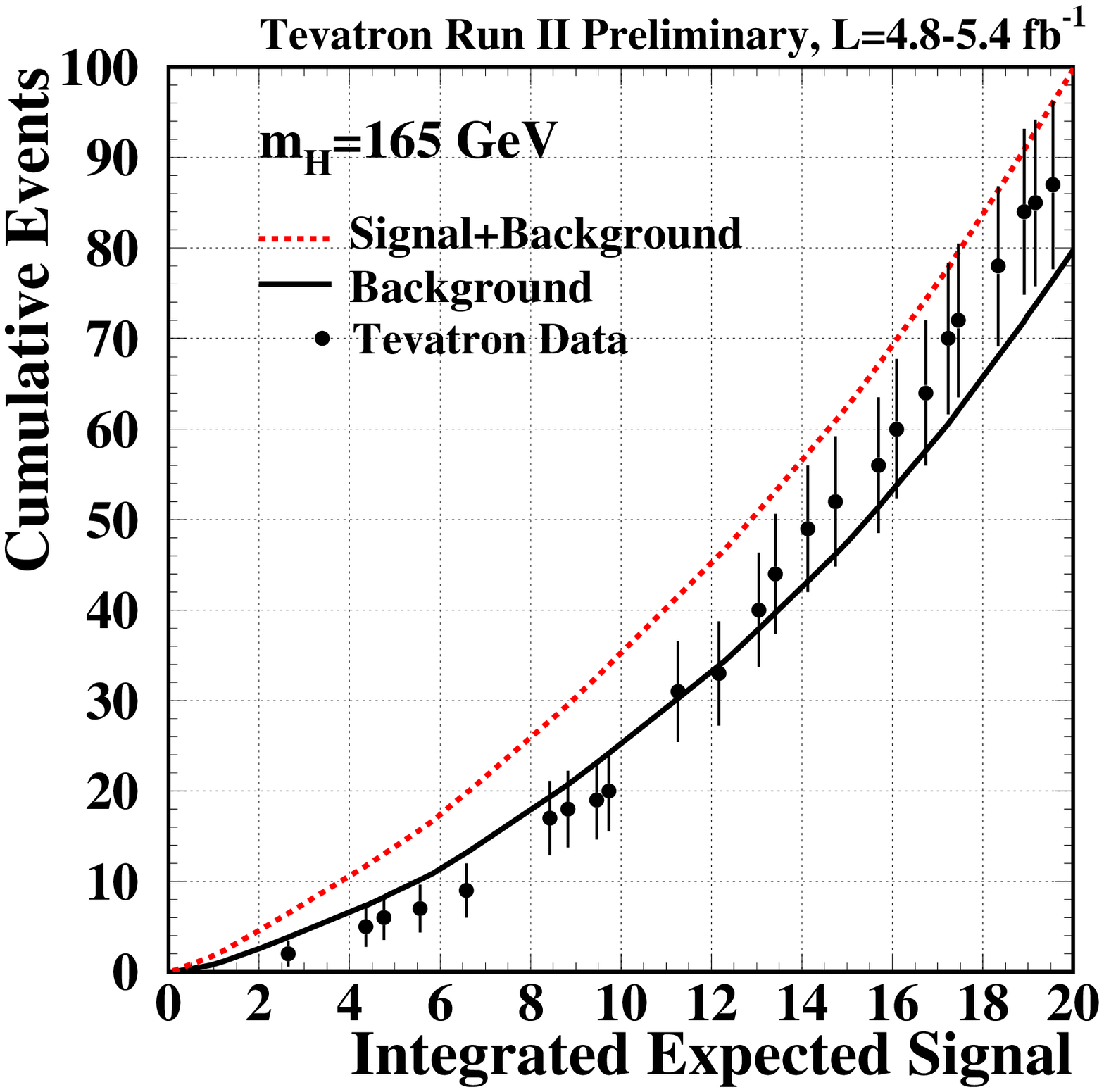}
 \caption{
 \label{fig:integ} Integrated distributions of $s/b$, starting at the high $s/b$ side.  The total signal+background
and background-only integrals are shown separately, along with the
data sums.  Data are only shown for bins that have 
data events in them.}
 \end{centering}
 \end{figure}

\section{Combining Channels} 

To gain confidence that the final result does not depend on the
details of the statistical formulation, 
we perform two types of combinations, using the
Bayesian and  Modified Frequentist approaches, which yield results that agree within 10\%.
Both methods rely on distributions in the final discriminants, and not just on
their single integrated values.  Systematic uncertainties enter on the
predicted number of signal and background events as well
as on the distribution of the discriminants in 
each analysis (``shape uncertainties'').
Both methods use likelihood calculations based on Poisson
probabilities.

\subsection{Bayesian Method}

Because there is no experimental information on the production cross section for
the Higgs boson, in the Bayesian technique~\cite{CDFHiggs} we assign a flat prior
for the total number of selected Higgs events.  For a given Higgs boson mass, the
combined likelihood is a product of likelihoods for the individual
channels, each of which is a product over histogram bins:

\begin{equation}
{\cal{L}}(R,{\vec{s}},{\vec{b}}|{\vec{n}},{\vec{\theta}})\times\pi({\vec{\theta}})
= \prod_{i=1}^{N_C}\prod_{j=1}^{Nbins} \mu_{ij}^{n_{ij}} e^{-\mu_{ij}}/n_{ij}!
\times\prod_{k=1}^{n_{np}}e^{-\theta_k^2/2}
\end{equation}

\noindent where the first product is over the number of channels
($N_C$), and the second product is over histogram bins containing
$n_{ij}$ events, binned in  ranges of the final discriminants used for
individual analyses, such as the dijet mass, neural-network outputs, 
or matrix-element likelihoods.
 The parameters that contribute to the
expected bin contents are $\mu_{ij} =R \times s_{ij}({\vec{\theta}}) + b_{ij}({\vec{\theta}})$ 
for the
channel $i$ and the histogram bin $j$, where $s_{ij}$ and $b_{ij}$ 
represent the expected background and signal in the bin, and $R$ is a scaling factor
applied to the signal to test the sensitivity level of the experiment.  
Truncated Gaussian priors are used for each of the nuisance parameters                                               
$\theta_k$, which define
the
sensitivity of the predicted signal and background estimates to systematic uncertainties.
These
can take the form of uncertainties on overall rates, as well as the shapes of the distributions
used for combination.   These systematic uncertainties can be far larger
than the expected SM signal, and are therefore important in the calculation of limits. 
The truncation
is applied so that no prediction of any signal or background in any bin is negative.
The posterior density function is
then integrated over all parameters (including correlations) except for $R$,
and a 95\% credibility level upper limit on $R$ is estimated
by calculating the value of $R$ that corresponds to 95\% of the area
of the resulting distribution.

\subsection{Modified Frequentist Method}

The Modified Frequentist technique relies on the $CL_s$ method, using
a log-likelihood ratio (LLR) as test statistic~\cite{DZHiggs}:
\begin{equation}
LLR = -2\ln\frac{p({\mathrm{data}}|H_1)}{p({\mathrm{data}}|H_0)},
\end{equation}
where $H_1$ denotes the test hypothesis, which admits the presence of
SM backgrounds and a Higgs boson signal, while $H_0$ is the null
hypothesis, for only SM backgrounds.  The probabilities $p$ are
computed using the best-fit values of the nuisance parameters for each
pseudo-experiment, separately for each of the two hypotheses, and include the
Poisson probabilities of observing the data multiplied by Gaussian
priors for the values of the nuisance parameters.  This technique
extends the LEP procedure~\cite{pdgstats} which does not involve a
fit, in order to yield better sensitivity when expected signals are
small and systematic uncertainties on backgrounds are
large~\cite{pflh}.

The $CL_s$ technique involves computing two $p$-values, $CL_{s+b}$ and $CL_b$.
The latter is defined by
\begin{equation}
1-CL_b = p(LLR\le LLR_{\mathrm{obs}} | H_0),
\end{equation}
where $LLR_{\mathrm{obs}}$ is the value of the test statistic computed for the
data. $1-CL_b$ is the probability of observing a signal-plus-background-like outcome 
without the presence of signal, i.e. the probability
that an upward fluctuation of the background provides  a signal-plus-background-like
response as observed in data.
The other $p$-value is defined by
\begin{equation}
CL_{s+b} = p(LLR\ge LLR_{\mathrm{obs}} | H_1),
\end{equation}
and this corresponds to the probability of a downward fluctuation of the sum
of signal and background in 
the data.  A small value of $CL_{s+b}$ reflects inconsistency with  $H_1$.
It is also possible to have a downward fluctuation in data even in the absence of
any signal, and a small value of $CL_{s+b}$ is possible even if the expected signal is
so small that it cannot be tested with the experiment.  To minimize the possibility
of  excluding  a signal to which there is insufficient sensitivity 
(an outcome  expected 5\% of the time at the 95\% C.L., for full coverage),
we use the quantity $CL_s=CL_{s+b}/CL_b$.  If $CL_s<0.05$ for a particular choice
of $H_1$, that hypothesis is deemed excluded at the 95\% C.L.

Systematic uncertainties are included  by fluctuating the predictions for
signal and background rates in each bin of each histogram in a correlated way when
generating the pseudo-experiments used to compute $CL_{s+b}$ and $CL_b$.

\subsection{Systematic Uncertainties} 

Systematic uncertainties differ
between experiments and analyses, and they affect the rates and shapes of the predicted
signal and background in correlated ways.  The combined results incorporate
the sensitivity of predictions to  values of nuisance parameters,
and include correlations between rates and shapes, between signals and backgrounds,
and between channels within experiments and between experiments.
More on these issues can be found in the
individual analysis notes~\cite{cdfWH2J} through~\cite{dzttH}.  Here we
consider only the largest contributions and correlations between and
within the two experiments.

\subsubsection{Correlated Systematics between CDF and D0}

The uncertainties on the measurements of the integrated luminosities are 6\%
(CDF) and 6.1\% (D0).  
Of these values, 4\% arises from the uncertainty
on the inelastic \pp~scattering cross section, which is correlated
between CDF and D0. 
CDF and D0 also share the assumed values and uncertainties on the production cross sections
for top-quark processes (\ttbar~and single top) and for electroweak processes
($WW$, $WZ$, and $ZZ$).  In order to provide a consistent combination, the values of these
cross sections assumed in each analysis are brought into agreement.  We use 
$\sigma_{t\bar{t}}=7.88\pm 0.79$~pb, following the calculation of Moch and Uwer~\cite{mochuwer}, assuming
a top quark mass $m_t=172.4\pm 1.2$~GeV/$c^2$~\cite{tevtop08}, 
and using the MRST2006nnlo PDF set~\cite{mrst2006}.  Other
calculations of $\sigma_{t\bar{t}}$ are similar~\cite{otherttbar}.  We use $\sigma_{\rm{SingleTop}}=3.38\pm0.34$~pb,
following the calculation of Kidonakis~\cite{kidonakis}.  Other calculations of $\sigma_{\rm{SingleTop}}$ are
similar for our purposes~\cite{harris}.  We use $\sigma_{WW}=12.4\pm 0.7$~pb, $\sigma_{WZ}=3.7 \pm 0.2$~pb,
and $\sigma_{ZZ}=3.8\pm 0.2$~pb, calculated with MCFM~\cite{mcfm}.

In many analyses, the dominant background yields are calibrated with data control samples.
The methods of measuring the multijet (``QCD'')
backgrounds differ between CDF and D0, and even between analyses within the collaborations, there is no
correlation assumed between these rates.  Similarly, the large
uncertainties on the background rates for $W$+heavy flavor (HF) and $Z$+heavy flavor
are considered at this time to be uncorrelated, as both CDF and D0 estimate
these rates using data control samples, but employ different techniques.
The calibrations of fake leptons, unvetoed $\gamma\rightarrow e^+e^-$ conversions,
$b$-tag efficiencies and mistag rates are performed by each collaboration
using independent data samples and methods, hence are considered uncorrelated.

\subsubsection{Correlated Systematic Uncertainties for CDF}
The dominant systematic uncertainties for the CDF analyses are shown
in Table~\ref{tab:cdfsystwhtdt} 
for the $W^\pm H\rightarrow W^\pm
b{\bar{b}}$ channels, in Table~\ref{tab:cdfvvbb1}
for the $(W,Z)H\rightarrow
\MET b{\bar{b}}$ channels, in Table~\ref{tab:cdfllbb1} 
for the $ZH\rightarrow
\ell^+\ell^-b{\bar{b}}$ channels, in Table~\ref{tab:cdfsystww0} for
the $H\rightarrow W^+W^-\rightarrow \ell^{\prime \pm}\nu \ell^{\prime
\mp}\nu$ channels, in Table~\ref{tab:cdfsysttautau} for the
$H\rightarrow\tau^+\tau^-$ channel, in Table~\ref{tab:cdfjjbbsyst} for
the $WH+ZH\rightarrow jjb{\bar{b}}$ channel, and in
Table~\ref{tab:cdfsystwww} for the $WH \rightarrow WWW
\rightarrow\ell^{\prime \pm}\ell^{\prime \pm}$ channel.  Each source
induces a correlated uncertainty across all CDF channels' signal and
background contributions which are sensitive to that source.  For
\hbb, the largest uncertainties on signal arise from a scale factor
for $b$-tagging (5.3-16\%), jet energy scale (1-20\%) and MC modeling
(2-10\%).  The shape dependence of the jet energy scale, $b$-tagging
and uncertainties on gluon radiation (``ISR'' and ``FSR'') are taken
into account for some analyses (see tables).  For \hww, the largest
uncertainty comes from MC modeling (5\%).  For simulated backgrounds,
the uncertainties on the expected rates range from 11-40\% (depending
on background).  The backgrounds with the largest systematic
uncertainties are in general quite small. Such uncertainties are
constrained by fits to the nuisance parameters, and they do not affect
the result significantly.  Because the largest background
contributions are measured using data, these uncertainties are treated
as uncorrelated for the \hbb~channels. For the \hww~channel, the
uncertainty on luminosity is taken to be correlated between signal and
background. The differences in the resulting limits when treating
the remaining uncertainties as correlated or uncorrelated, is less than $5\%$.

\subsubsection{Correlated Systematic Uncertainties for D0 }
The dominant systematic uncertainties for D0 analyses are shown in Tables 
V,\ref{tab:d0vvbb},\ref{tab:d0llbb1},\ref{tab:d0systwww},\ref{tab:d0systww}, 
\ref{tab:d0systtth}, and \ref{tab:d0systgg}.
Each source induces a correlated uncertainty across all D0 channels
sensitive to that source. Wherever appropriate the impact of the systematic effect on both the rate and
shape of the predicted signal and background is included.  
For the low mass, \hbb~analyses, the largest sources of uncertainty originate from
the $b$-tagging rate ($\sim$5-10\% per tagged jet), the determination of the jet energy, acceptance,
and resolution ($\sim$5-10\%), 
the normalization of the W and Z + heavy flavor backgrounds ($\sim$20\%) and the determination 
of the multijet background ($\sim 25\%$). 
For the \hww~and \www analyses, one of the largest uncertainties is associated with the lepton
measurement and acceptance, and is $\sim$5-10\%  depending on
the final state.  Significant sources for all analyses are
the uncertainties on the luminosity and the cross sections for the simulated backgrounds, and are
$\sim 6\%$ and 6-10\% respectively. 
All systematic uncertainties arising from the
same source are taken to be correlated between the different backgrounds and
between signal and background.

\begin{table}[t]
\caption{Systematic uncertainties on the signal and background contributions for
 CDF's $WH\rightarrow\ell\nu b{\bar{b}}$ tight (TDT), loose (LDT)
 double tag, looser (LDTX) double tag, and single tag (ST) channels.  Systematic uncertainties
 are listed by name; see the original references for a detailed
 explanation of their meaning and on how they are derived.  Systematic
 uncertainties for $WH$ shown in this table are obtained for $m_H=115$
 GeV/$c^2$.  Uncertainties are relative, in percent, and are symmetric
 unless otherwise indicated.  }
\label{tab:cdfsystwhtdt}
\vskip 0.1cm
{\centerline{CDF: tight (TDT) and loose (LDT) double-tag $WH\rightarrow\ell\nu b{\bar{b}}$ relative uncertainties (\%)}} 
\vskip 0.099cm
\begin{ruledtabular}
\begin{tabular}{lcccccc}\\
Contribution              & $W$+HF & Mistags & Top & Diboson & Non-$W$ & $WH$  \\ \hline
Luminosity ($\sigma_{\mathrm{inel}}(p{\bar{p}})$)          
                          & 0      & 0       & 3.8 & 3.8     & 0       &    3.8   \\
Luminosity Monitor        & 0      & 0       & 4.4 & 4.4     & 0       &    4.4   \\
Lepton ID                 & 0      & 0       & 2   & 2       & 0       &    2   \\
Jet Energy Scale          & 0      & 0       & 0   & 0       & 0       &    2   \\
Mistag Rate               & 0      & 35     & 0   & 0       & 0       &    0   \\
$B$-Tag Efficiency          & 0      & 0       & 8.6 & 8.6     & 0       &    8.6   \\
$t{\bar{t}}$ Cross Section  & 0    & 0       & 10  & 0       & 0       &    0   \\
Diboson Rate              & 0      & 0       & 0   & 11.5    & 0       &    0   \\
Signal Cross Section      & 0      & 0       & 0   & 0       & 0       &    5 \\
HF Fraction in W+jets     &    45  & 0       & 0   & 0       & 0       &    0   \\
ISR+FSR+PDF               & 0      & 0       & 0   & 0       & 0       &    5 \\ 
QCD Rate                  & 0      & 0       & 0   & 0       & 40      &    0   \\
\end{tabular}
\end{ruledtabular}
\vskip 0.3cm  
 {\centerline{CDF: looser double-tag (LDTX) $WH\rightarrow\ell\nu b{\bar{b}}$ relative uncertainties (\%)}}
\vskip 0.099cm
\label{tab:cdfsystwhldt}
\begin{ruledtabular}
\begin{tabular}{lcccccc}\\
Contribution              & $W$+HF & Mistags & Top & Diboson & Non-$W$ & $WH$  \\ \hline
Luminosity ($\sigma_{\mathrm{inel}}(p{\bar{p}})$)          
                          & 0      & 0       & 3.8 & 3.8     & 0       &    3.8   \\
Luminosity Monitor        & 0      & 0       & 4.4 & 4.4     & 0       &    4.4   \\
Lepton ID                 & 0      & 0       & 2   & 2       & 0       &    2   \\
Jet Energy Scale          & 0      & 0       & 0   & 0       & 0       &    2.2   \\
Mistag Rate               & 0      & 36     & 0   & 0       & 0       &    0   \\
$B$-Tag Efficiency          & 0      & 0       & 13.6 & 13.6     & 0       &    13.6   \\
$t{\bar{t}}$ Cross Section  & 0    & 0       & 10  & 0       & 0       &    0   \\
Diboson Rate              & 0      & 0       & 0   & 11.5    & 0       &    0   \\
Signal Cross Section      & 0      & 0       & 0   & 0       & 0       &    5 \\
HF Fraction in W+jets     &    45  & 0       & 0   & 0       & 0       &    0   \\
ISR+FSR+PDF               & 0      & 0       & 0   & 0       & 0       &    7.7 \\ 
QCD Rate                  & 0      & 0       & 0   & 0       & 40      &    0   \\
\end{tabular}
\end{ruledtabular}
\vskip 0.3cm                                                                            
{\centerline{CDF: single tag (ST) $WH\rightarrow\ell\nu b{\bar{b}}$ relative uncertainties (\%)}}
\vskip 0.099cm
\label{tab:cdfsystwhst}
\begin{ruledtabular}
\begin{tabular}{lcccccc}\\
Contribution              & $W$+HF & Mistags & Top & Diboson & Non-$W$ & $WH$  \\ \hline
Luminosity ($\sigma_{\mathrm{inel}}(p{\bar{p}})$)          
                          & 0      & 0       & 3.8 & 3.8     & 0       &    3.8   \\
Luminosity Monitor        & 0      & 0       & 4.4 & 4.4     & 0       &    4.4   \\
Lepton ID                 & 0      & 0       & 2   & 2       & 0       &    2   \\
Jet Energy Scale          & 0      & 0       & 0   & 0       & 0       &    2   \\
Mistag Rate               & 0      & 35    & 0   & 0       & 0       &    0   \\
$B$-Tag Efficiency          & 0      & 0       & 4.3 & 4.3     & 0       &    4.3   \\
$t{\bar{t}}$ Cross Section  & 0    & 0       & 10  & 0       & 0       &    0   \\
Diboson Rate              & 0      & 0       & 0   & 11.5    & 0       &    0   \\
Signal Cross Section      & 0      & 0       & 0   & 0       & 0       &    5 \\
HF Fraction in W+jets     &    42  & 0       & 0   & 0       & 0       &    0   \\
ISR+FSR+PDF               & 0      & 0       & 0   & 0       & 0       &    3.0 \\ 
QCD Rate                  & 0      & 0       & 0   & 0       & 40      &    0   \\
\end{tabular}
\end{ruledtabular}
\end{table}

\clearpage


\begin{table}[h]
\label{tab:d0systwh1}
\caption{Systematic uncertainties on the signal contributions  for D0's
$WH\rightarrow\ell\nu b{\bar{b}}$ single (ST) and double tag (DT) channels.
%
%
Systematic uncertainties are listed by name, see the original 
references for a detailed explanation of their meaning and on how they are derived.  
Systematic uncertainties for $WH$ shown in this table are obtained for $m_H=115$ GeV/$c^2$.
  Uncertainties are
relative, in percent, and are symmetric unless otherwise indicated.  }
\vskip 0.2cm
{\centerline{D0: Single Tag (ST) $WH \rightarrow\ell\nu b\bar{b}$ analysis relative uncertainties (\%)}}
\vskip 0.099cm
\begin{ruledtabular}
\begin{tabular}{l c c c c c c c }\\
Contribution  &~WZ/WW~&Wbb/Wcc&Wjj/Wcj&$~~~t\bar{t}~~~$&single top&Multijet& ~~~WH~~~\\ 
\hline                                                                            
Luminosity                &  6    &  6    &  6    &  6    &  6    &  0    &  6    \\ 
Trigger eff.              &  2--5 &  2--5 &  2--5 &  2--5 &  2--5 &  0    &  2--5 \\       
EM ID/Reco eff./resol.    &     3 &     3 &     3 &     3 &     3 &  0    &     3 \\       
Muon ID/Reco eff./resol.  &   4.1 &   4.1 &   4.1 &   4.1 &   4.1 &  0    &   4.1 \\        
Jet ID/Reco eff.          &     2 &     2 &     2 &     2 &     2 &  0    &     2 \\ 
Jet Energy Scale          &     3 &     3 &     3 &     3 &     3 &  0    &     3 \\       
Jet mult./frag./modeling  &   3.5 &   3.5 &   3.5 &   3.5 &   3.5 &  0    &   3.5 \\       
$b$-tagging/taggability   &     4 &     4 &    11 &     4 &     4 &  0    &     4 \\ 
Cross Section             &     6 &     9 &     9 &    10 &    10 &  0    &     6 \\       
Heavy-Flavor K-factor     &  0    &    20 &     0 &  0    &  0    &  0    &  0    \\       
Instrumental-WH           &  0    &  0    &  0   &  0    &  0    &    26 &  0    \\ 
\end{tabular}                                                                            
\end{ruledtabular}
\vskip 0.5cm                                                                             
{\centerline{D0: Double Tag (DT) $WH \rightarrow\ell\nu b\bar{b}$ analysis relative uncertainties (\%)}}
\vskip 0.099cm
\begin{ruledtabular}
\begin{tabular}{ l c c c c c c c }   \\                                             
Contribution  &~WZ/WW~&Wbb/Wcc&Wjj/Wcj&$~~~t\bar{t}~~~$&single top&Multijet& ~~~WH~~~\\ 
\hline                                                                            
Luminosity                &  6    &  6    &  6    &  6    &  6    &  0    &  6    \\ 
Trigger eff.              &  2--5 &  2--5 &  2--5 &  2--5 &  2--5 &  0    &  2--5 \\       
EM ID/Reco eff./resol.    &     3 &     3 &     3 &     3 &     3 &  0    &     3 \\       
Muon ID/Reco eff./resol.  &   4.1 &   4.1 &   4.1 &   4.1 &   4.1 &  0    &   4.1 \\        
Jet ID/Reco eff.          &     2 &     2 &     2 &     2 &     2 &  0    &     2 \\ 
Jet Energy Scale          &     3 &     3 &     3 &     3 &     3 &  0    &     3 \\       
Jet mult./frag./modeling  &   3.5 &   3.5 &   3.5 &   3.5 &   3.5 &  0    &   3.5 \\       
$b$-tagging/taggability   &     6 &     6 &    20 &     6 &     6 &  0    &     6 \\ 
Cross Section             &     6 &     9 &     9 &    10 &    10 &  0    &     6 \\       
Heavy-Flavor K-factor     &  0    &    20 &     0 &  0    &  0    &  0    &  0    \\       
Instrumental-WH           &  0    &  0    &  0    &  0    &  0    &    26 &  0    \\ 
\end{tabular}                                                                                           
\end{ruledtabular}
%
%


\end{table}


\begin{table}
\caption{Systematic uncertainties for CDF's $WH,ZH\rightarrow\MET b{\bar{b}}$ tight (TDT and loose (LDT) double-tag, and single-tag (ST) channels.
Systematic uncertainties are listed by name; see the original references for a detailed explanation of their meaning and on how they are derived.  
Systematic uncertainties for $ZH$ and $WH$ shown in this table are obtained for $m_H=120$~GeV/$c^2$.
Uncertainties are relative, in percent, and are symmetric unless otherwise indicated.  }
\label{tab:cdfvvbb1}
\vskip 0.5cm                                                                                                          
{\centerline{CDF: $WH,ZH\rightarrow\MET b{\bar{b}}$ tight double-tag (TDT) channel relative uncertainties (\%)}}
\vskip 0.099cm                                                                                                          
\begin{footnotesize}
\begin{ruledtabular}
      \begin{tabular}{lcccccccc}\\
                          & ZH & WH &Multijet& Top Pair & S. Top & Di-boson  & W + h.f.  & Z + h.f. \\\hline
        \multicolumn{9}{l}{\it{Correlated uncertainties}}                                       \\\hline
        Luminosity       & 3.8 & 3.8 &       & 3.8 & 3.8 & 3.8     & 3.8     & 3.8     \\
        Lumi Monitor      & 4.4 & 4.4 &       & 4.4 & 4.4 & 4.4     & 4.4     & 4.4     \\
        Tagging SF        & 8.6 & 8.6 &       & 8.6 & 8.6 & 8.6     & 8.6     & 8.6     \\
      Trigger Eff. (shape)& 1.0 & 1.2 & 1.1 & 0.7 & 1.1 & 1.6     & 1.7     & 1.3     \\
        Lepton Veto       & 2.0 & 2.0 &       & 2.0 & 2.0 &2.0      & 2.0     & 2.0     \\
        PDF Acceptance    & 2.0 & 2.0 &       & 2.0 & 2.0 &2.0      & 2.0     & 2.0     \\
        JES (shape)       & $^{+3.0}_{-3.0}$ 
                                  & $^{+3.5}_{-4.7}$ 
                                          &  $^{-4.0}_{+3.8}$ 
                                                  & $^{+1.1}_{-1.1}$ 
                                                          & $^{+2.4}_{-4.7}$ 
                                                                  & $^{+8.2}_{-6.1}$ 
                                                                             & $^{+7.3}_{-11.8}$  
                                                                                          & $^{+6.5}_{-8.3}$    \\
        ISR               & \multicolumn{2}{c}{$^{+4.4}_{+3.7}$} &       &       &       &           &           &      \\
        FSR               & \multicolumn{2}{c}{$^{+1.8}_{+4.4}$} &       &       &       &           &           &      \\\hline
        \multicolumn{9}{l}{\it{Uncorrelated uncertainties}}                             \\\hline
        Cross-Section     &  5  & 5 &       & 10 & 10 & 11.5    & 40      & 40      \\
        Multijet Norm.  (shape)   &       & & 17 &       & &          &           &           \\
      \end{tabular}
\end{ruledtabular}
\end{footnotesize}
\vskip 0.5cm                                                                                                          
{\centerline{CDF: $WH,ZH\rightarrow\MET b{\bar{b}}$ loose double-tag (LDT) channel relative uncertainties (\%)}}
\vskip 0.099cm                                                                                                          
\label{tab:cdfvvbb2} 
\begin{footnotesize}
 \begin{ruledtabular}
     \begin{tabular}{lcccccccc}\\
                          & ZH & WH & Multijet & Top Pair & S. Top  & Di-boson  & W + h.f.  & Z + h.f. \\\hline
        \multicolumn{9}{l}{\it{Correlated uncertainties}}                                       \\\hline
        Luminosity       & 3.8  & 3.8  &     & 3.8  & 3.8  & 3.8      & 3.8      & 3.8     \\
        Lumi Monitor      & 4.4  & 4.4  &     & 4.4  & 4.4  & 4.4      & 4.4      & 4.4     \\
        Tagging SF        & 9.1 & 9.1 &     & 9.1 & 9.1 & 9.1     & 9.1     & 9.1     \\
     Trigger Eff. (shape) & 1.2  & 1.3  &1.1& 0.7  & 1.2  & 1.2      & 1.8      & 1.3     \\
        Lepton Veto       & 2.0  & 2.0  &     & 2.0  & 2.0  &2.0       & 2.0      & 2.0     \\
        PDF Acceptance    & 2.0  & 2.0  &     & 2.0  & 2.0  &2.0       & 2.0      & 2.0     \\
        JES (shape)       & $^{+3.7}_{-3.7}$ 
                                   & $^{+4.0}_{-4.0}$ 
                                            & $^{-5.4}_{+5.2}$     
                                                  & $^{+1.1}_{-0.7}$ 
                                                           & $^{+4.2}_{-4.2}$ 
                                                                    & $^{+7.0}_{-7.0}$ 
                                                                                 & $^{+1.3}_{-7.6}$  
                                                                                              & $^{+6.2}_{-7.1}$    \\
        ISR               & \multicolumn{2}{c}{$^{+1.4}_{-2.9}$} &       &       &       &           &           &      \\
        FSR               & \multicolumn{2}{c}{$^{+5.3}_{+2.5}$} &       &       &       &           &           &      \\\hline
        \multicolumn{9}{l}{\it{Uncorrelated uncertainties}}                             \\\hline
        Cross-Section     &  5.0   & 5.0 &      & 10 & 10 & 11.5    & 40      & 40      \\
        Multijet Norm.  (shape)   &       & & 11 &       & &          &           &           \\\hline
      \end{tabular}
\end{ruledtabular}
\end{footnotesize}
\vskip 0.5cm                                                                                                          
{\centerline{CDF: $WH,ZH\rightarrow\MET b{\bar{b}}$ single-tag (ST) channel relative uncertainties (\%)}}
\vskip 0.099cm                                                                                                          
\label{tab:cdfvvbb3}
\begin{footnotesize}
\begin{ruledtabular}
      \begin{tabular}{lcccccccc}\\
                          & ZH & WH & Multijet & Top Pair & S. Top  & Di-boson  & W + h.f.  & Z + h.f. \\\hline
        \multicolumn{9}{l}{\it{Correlated uncertainties}}                                       \\\hline
        Luminosity       & 3.8  & 3.8  &     & 3.8  & 3.8  & 3.8      & 3.8      & 3.8     \\
        Lumi Monitor      & 4.4  & 4.4  &     & 4.4  & 4.4  & 4.4      & 4.4      & 4.4     \\
        Tagging SF        & 4.3  & 4.3  &     & 4.3  & 4.3  & 4.3      & 4.3      & 4.3     \\
     Trigger Eff. (shape) & 0.9  & 1.1  &1.1& 0.7  & 1.1  & 1.3      & 2.0      & 1.4     \\
        Lepton Veto       & 2.0  & 2.0  &     & 2.0  & 2.0  &2.0       & 2.0      & 2.0     \\
        PDF Acceptance    & 2.0  & 2.0  &     & 2.0  & 2.0  &2.0       & 2.0      & 2.0     \\
        JES (shape)       & $^{+3.8}_{-3.8}$ 
                                   & $^{+3.8}_{-3.8}$ 
                                            & $^{-5.2}_{+5.6}$     
                                                  & $^{+0.7}_{-0.8}$ 
                                                           & $^{+4.6}_{-4.6}$ 
                                                                    & $^{+7.0}_{-5.6}$ 
                                                                                 & $^{+12.4}_{-12.7}$  
                                                                                              & $^{+8.3}_{-8.1}$    \\
        ISR               & \multicolumn{2}{c}{$^{-1.0}_{-1.5}$} &       &       &       &           &           &      \\
        FSR               & \multicolumn{2}{c}{$^{+2.0}_{-0.1}$} &       &       &       &           &           &      \\\hline
        \multicolumn{9}{l}{\it{Uncorrelated uncertainties}}                             \\\hline
        Cross-Section     &  5.0   & 5.0 &      & 10 & 10 & 11.5    & 40      & 40      \\
        Multijet Norm.  (shape)   &       & & 3.9 &       & &          &           &           \\\hline
      \end{tabular}
\end{ruledtabular}
\end{footnotesize}
\end{table}

\begin{table}
\begin{center}
\caption{Systematic uncertainties on the contributions for D0's $ZH\rightarrow \nu \nu b{\bar{b}}$ single (ST) 
 and tight-loose double-tag (TLDT) channels.
Systematic uncertainties are listed by name; see the original references for a detailed explanation of their meaning 
and on how they are derived.  
Systematic uncertainties for $ZH$, $WH$  shown in this table are obtained for $m_H=115$ GeV/$c^2$.
Uncertainties are
relative, in percent, and are symmetric unless otherwise indicated. Shape uncertainties are labeledwith an ``s''. }
\label{tab:d0vvbb}
\vskip 0.5cm                                                                                                          
{\centerline{D0: Single Tag (ST)~ $ZH \rightarrow \nu\nu b \bar{b}$ analysis relative uncertainties (\%)}}
\vskip 0.099cm                                                                                                      
\begin{ruledtabular}
\begin{tabular}{ l c c c c c c }
Contribution                          &~WZ/ZZ~        &~Z+jets~      &~W+jets~      &~~~$t\bar{t}$    &~~ZH,WH~~\\ \hline
Jet Energy Scale pos/neg (S)          &  6.9/-5.2     &  6.8/-5.5    &  6/-4.8      &  -1.4/1.2       &  2.2/-3.4      \\
Jet ID (S)                            &  0.8          &  1.0         &  1.0         &  0.3            &  0.8      \\
Jet Resolution pos/neg (S)            &  5.0/1.8      &  5.0/-5.5    &  3.3/-1.0    &  -0.8/0.9       &  -0.8/-0.1      \\
MC Heavy flavor $b$-tagging pos/neg (S) &  4.2/-4.4   &  4.0/-4.4    &  3.9/-4.2    &  3.8/-4.3       &  0.9/-2.0      \\
MC light flavor $b$-tagging pos/neg (S) &  3.2/-3.3   &  0.3/-0.3    &  0.5/-0.5    &  0.0            &  0.0      \\
Direct taggability (S)                &  5.6          &  3.2         &  3.1         &  0.9            &  3.7      \\
Vertex Confirmation (S)               &  0.6          &  3.1         &  3.2         &  0.5            &  2.2      \\
Trigger efficiency (S)                      &  3.5          &  3.3         &  3.3         &  3.2            &  3.4      \\
ALPGEN MLM pos/neg(S)                 &  -            &  0.4/0.0     &  0.5/0.0     &  -              &  -      \\
ALPGEN Scale (S)                      &  -            &  0.6         &  0.6         &  -              &  -      \\
Underlying Event (S)                  &  -            &  0.4         &  0.4         &  -              &  -      \\
Parton Distribution Function (S)      &  2.0          &  2.0         &  2.0         &  2.0            &  2.0      \\
EM ID                                 &  0.2          &  0           &  0.3         &  0.1            &  0.2      \\
Muon ID                               &  1.1          &  0.3         &  1.8         &  0.9            &  0.9      \\
Cross Section                         &  7            &  6.0         &  6.0         &  10             &  6.0      \\
Heavy Flavor Ratio                   &  -            &  20          &  20          &  -              &  -      \\
Luminosity                            &  6.1          &  6.1         &  6.1         &  6.1            &  6.1  \\
\end{tabular}
\end{ruledtabular}
         
\vskip 0.5cm                                                                                                          
{\centerline{D0: Double Tag (TLDT)~ $ZH \rightarrow \nu\nu b \bar{b}$ analysis relative uncertainties (\%)}}
\vskip 0.099cm                                                                                                          
\label{tab:d0vvbb2}
\begin{ruledtabular}
\begin{tabular}{ l  c  c  c  c  c  c }
Contribution                          &~WZ/ZZ~        &~Z+jets~      &~W+jets~      &~~~$t\bar{t}$    &~~ZH,WH~~\\ \hline
Jet Energy Scale pos/neg (S)          &  9.3/-11.1    &  4.1/-5.6    &  7.1/-5.2    &  -0.9/0.5       &  2.3/-3.0      \\
Jet ID (S)                            &  1.7          &  0.1         &  1.1         &  0.0            &  0.8      \\
Jet Resolution pos/neg  (S)           &  -0.3/-7.4    &  1.2/-3.2    &  1.3         &  -1.0/0.5       &  -1.2/0.7      \\
MC Heavy flavor $b$-tagging pos/neg (S) &  7.6/-7.4   &  7.9/-7.7    &  7.8/-7.6    &  8.4/-8.2       &  8.5/-8.4      \\
MC light flavor $b$-tagging pos/neg (S) &  2.1/-2.1   &  0.7/-0.7    &  0.9/-0.9    &  0.5/-0.5       &  0.1/-0.1      \\
Direct taggability (S)                &  4.5          &  5.5         &  4.5         &  1.6            &  3.7      \\
Vertex Confirmation (S)               &  2.4          &  0.0         &  4.6         &  2.4            &  2.5      \\
Trigger efficiency (S)                      &  3.4          &  3.3         &  3.3         &  3.4            &  3.4      \\
ALPGEN MLM pos/neg (S)                &  -            &  0.0/0.3     &  0.6/-0.1    &  -              &  -      \\
ALPGEN Scale (S)                      &  -            &  0.4         &  0.8         &  -              &  -      \\
Underlying Event (S)                  &  -            &  0.4         &  0.4         &  -              &  -      \\
Parton Distribution Function (S)      &  2.0          &  2.0         &  2.0         &  2.0            &  2.0      \\
EM ID                                 &  0.2          &  0           &  0.4         &  0.7            &  0.1      \\
Muon ID                               &  0.9          &  0.5         &  1.0         &  1.9            &  1.0      \\
Cross Section                         &  7            &  6.0         &  6.0         & 10              &  6.0      \\
Heavy Flavor Ratio                   &  -            &  20          &  20          &  -              &  -      \\
Luminosity                            &  6.1          &  6.1         &  6.1         &  6.1            &  6.1  \\
\end{tabular}
\end{ruledtabular}
\end{center}
\end{table}

%
\begin{table}
\begin{center}
\caption{Systematic uncertainties on the contributions for CDF's $ZH\rightarrow \ell^+\ell^-b{\bar{b}}$ single-tag (ST), tight double-tag
(TDT), and loose double-tag (LDT) channels.  The channels are further divided into low- and high $s/b$ categories.
Systematic uncertainties are listed by name; see the original references for a detailed explanation of their meaning and on how they are derived.  
Systematic uncertainties for $ZH$  shown in this table are obtained for $m_H=115$ GeV/$c^2$.
Uncertainties are relative, in percent, and are symmetric unless otherwise indicated. }
\label{tab:cdfllbb1}
\vskip 0.3cm                                                                                                          
{\centerline{CDF: Single Tag High $s/b$ (ST High)~ $ZH \rightarrow \ell\ell b \bar{b}$ analysis relative uncertainties (\%)}}
\vskip 0.099cm                                                                                                          
\begin{tabular}{|l|c|c|c|c|c|c|c|c|} \hline
Contribution   & ~Fakes~ & ~~~Top~~~  & ~~$WZ$~~ & ~~$ZZ$~~ & ~$Z+b{\bar{b}}$~ & ~$Z+c{\bar{c}}$~& ~$Z+$mistag~ & ~~~$ZH$~~~ \\ \hline
Luminosity ($\sigma_{\mathrm{inel}}(p{\bar{p}})$)          & 0     &    3.8 &    3.8 &    3.8 &    3.8           &    3.8          & 0        &    3.8  \\
Luminosity Monitor        & 0     &    4.4 &    4.4 &    4.4 &    4.4           &    4.4          & 0        &    4.4  \\
Lepton ID    & 0     &    1 &    1 &    1 &    1           &    1          & 0        &    1  \\
Lepton Energy Scale    & 0     &    1.5 &    1.5 &    1.5 &    1.5           &    1.5          & 0        &    1.5  \\
$ZH$ Cross Section    & 0     &    0 &    0 &    0 &    0           &    0          & 0        &    5 \\
Fake Leptons       & 50    & 0    & 0    & 0    & 0              & 0             & 0        & 0     \\
Jet Energy Scale  (shape dep.)       & 0     & 
  $^{+1.9}_{-2.2}$   & 
  $^{+3.1}_{-4.6}$   & 
  $^{+3.5}_{-5.1}$   & 
  $^{+10.6}_{-9.6}$   & 
  $^{+9.5}_{-9.4}$   & 
  0   & 
  $^{+2.6}_{-2.2}$   \\ 
Mistag Rate (shape dep.)      & 0     & 0    & 0    & 0    & 0              & 0             &   $^{+14.7}_{-14.8}$     & 0     \\
B-Tag Efficiency      & 0     &    4 &    4 &    4 &    4           &   4          & 0        &    4  \\
$t{\bar{t}}$ Cross Section         & 0     &   20 & 0    & 0    & 0              & 0             & 0        & 0     \\
Diboson Cross Section        & 0     & 0    & 11.5   & 11.5    & 0              & 0             & 0        & 0     \\
$\sigma(p{\bar{p}}\rightarrow Z+HF)$      & 0     & 0    & 0    & 0    &  40            & 40           & 0        & 0     \\
ISR (shape dep.)           & 0     & 0    & 0    & 0    & 0              & 0             & 0        &   $^{-3.2}_{-4.2}$     \\
FSR (shape dep.)           & 0     & 0    & 0    & 0    & 0              & 0             & 0        &   $^{-0.01}_{-1.3}$     \\
\hline
\end{tabular}
\vskip 0.3cm                                                                                                          
{\centerline{CDF: Single Tag Low $s/b$ (ST Low)~ $ZH \rightarrow \ell\ell b \bar{b}$ analysis relative uncertainties (\%)}}
\vskip 0.099cm                                                                                                          
\begin{tabular}{|l|c|c|c|c|c|c|c|c|} \hline
Contribution   & ~Fakes~ & ~~~Top~~~  & ~~$WZ$~~ & ~~$ZZ$~~ & ~$Z+b{\bar{b}}$~ & ~$Z+c{\bar{c}}$~& ~$Z+$mistag~ & ~~~$ZH$~~~ \\ \hline
Luminosity ($\sigma_{\mathrm{inel}}(p{\bar{p}})$)          & 0     &    3.8 &    3.8 &    3.8 &    3.8           &    3.8          & 0        &    3.8  \\
Luminosity Monitor        & 0     &    4.4 &    4.4 &    4.4 &    4.4           &    4.4          & 0        &    4.4  \\
Lepton ID    & 0     &    1 &    1 &    1 &    1           &    1          & 0        &    1  \\
Lepton Energy Scale    & 0     &    1.5 &    1.5 &    1.5 &    1.5           &    1.5          & 0        &    1.5  \\
$ZH$ Cross Section    & 0     &    0 &    0 &    0 &    0           &    0          & 0        &    5 \\
Fake Leptons       & 50    & 0    & 0    & 0    & 0              & 0             & 0        & 0     \\
Jet Energy Scale  (shape dep.)       & 0     & 
  $^{+1.8}_{-1.6}$   & 
  $^{+7.0}_{-4.5}$   & 
  $^{+2.7}_{-6.3}$   & 
  $^{+11.7}_{-10.2}$   & 
  $^{+10.0}_{-10.2}$   & 
  0   & 
  $^{+7.4}_{+1.6}$   \\ 
Mistag Rate (shape dep.)      & 0     & 0    & 0    & 0    & 0              & 0             &   $^{+14.8}_{-14.9}$     & 0     \\
B-Tag Efficiency      & 0     &    4 &    4 &    4 &    4           &   4          & 0        &    4  \\
$t{\bar{t}}$ Cross Section         & 0     &   20 & 0    & 0    & 0              & 0             & 0        & 0     \\
Diboson Cross Section        & 0     & 0    & 11.5   & 11.5    & 0              & 0             & 0        & 0     \\
$\sigma(p{\bar{p}}\rightarrow Z+HF)$      & 0     & 0    & 0    & 0    &  40            & 40           & 0        & 0     \\
ISR (shape dep.)           & 0     & 0    & 0    & 0    & 0              & 0             & 0        &   $^{+12.5}_{+3.3}$     \\
FSR (shape dep.)           & 0     & 0    & 0    & 0    & 0              & 0             & 0        &   $^{+9.0}_{+6.3}$     \\
\hline
\end{tabular}
\vskip 0.3cm   
{\centerline{CDF: Tight Double Tag High $s/b$ (TDT High)~ $ZH \rightarrow \ell\ell b \bar{b}$ analysis relative uncertainties (\%)}}
\vskip 0.099cm                                                                                                          
\begin{tabular}{|l|c|c|c|c|c|c|c|c|} \hline
Contribution   & ~Fakes~ & ~~~Top~~~  & ~~$WZ$~~ & ~~$ZZ$~~ & ~$Z+b{\bar{b}}$~ & ~$Z+c{\bar{c}}$~& ~$Z+$mistag~ & ~~~$ZH$~~~ \\ \hline
Luminosity ($\sigma_{\mathrm{inel}}(p{\bar{p}})$)          & 0     &    3.8 &    3.8 &    3.8 &    3.8           &    3.8          & 0        &    3.8  \\
Luminosity Monitor        & 0     &    4.4 &    4.4 &    4.4 &    4.4           &    4.4          & 0        &    4.4  \\
Lepton ID    & 0     &    1 &    1 &    1 &    1           &    1          & 0        &    1  \\
Lepton Energy Scale    & 0     &    1.5 &    1.5 &    1.5 &    1.5           &    1.5          & 0        &    1.5  \\
$ZH$ Cross Section    & 0     &    0 &    0 &    0 &    0           &    0          & 0        &    5 \\
Fake Leptons       & 50    & 0    & 0    & 0    & 0              & 0             & 0        & 0     \\
Jet Energy Scale  (shape dep.)       & 0     & 
  $^{+1.6}_{-1.1}$   & 
  $^{+0.0}_{-0.0}$   & 
  $^{+1.8}_{-2.7}$   & 
  $^{+5.9}_{-6.8}$   & 
  $^{+6.0}_{-5.9}$   & 
  0   & 
  $^{+2.0}_{+0.01}$   \\ 
Mistag Rate (shape dep.)      & 0     & 0    & 0    & 0    & 0              & 0             &   $^{+30.7}_{-26.6}$     & 0     \\
B-Tag Efficiency      & 0     &    8 &    8 &    8 &    8           &   8          & 0        &    8  \\
$t{\bar{t}}$ Cross Section         & 0     &   20 & 0    & 0    & 0              & 0             & 0        & 0     \\
Diboson Cross Section        & 0     & 0    & 11.5   & 11.5    & 0              & 0             & 0        & 0     \\
$\sigma(p{\bar{p}}\rightarrow Z+HF)$      & 0     & 0    & 0    & 0    &  40            & 40           & 0        & 0     \\
ISR (shape dep.)           & 0     & 0    & 0    & 0    & 0              & 0             & 0        &   $^{-2.0}_{+1.2}$     \\
FSR (shape dep.)           & 0     & 0    & 0    & 0    & 0              & 0             & 0        &   $^{-0.01}_{+0.01}$     \\
\hline
\end{tabular}
\end{center}
\end{table}

\begin{table}
\begin{center}
\label{tab:cdfllbb2}
\vskip 0.8cm      

{\centerline{CDF: Tight Double Tag Low $s/b$ (TDT Low)~ $ZH \rightarrow \ell\ell b \bar{b}$ analysis relative uncertainties (\%)}}
\vskip 0.099cm                                                                                                          
\begin{tabular}{|l|c|c|c|c|c|c|c|c|} \hline
Contribution   & ~Fakes~ & ~~~Top~~~  & ~~$WZ$~~ & ~~$ZZ$~~ & ~$Z+b{\bar{b}}$~ & ~$Z+c{\bar{c}}$~& ~$Z+$mistag~ & ~~~$ZH$~~~ \\ \hline
Luminosity ($\sigma_{\mathrm{inel}}(p{\bar{p}})$)          & 0     &    3.8 &    3.8 &    3.8 &    3.8           &    3.8          & 0        &    3.8  \\
Luminosity Monitor        & 0     &    4.4 &    4.4 &    4.4 &    4.4           &    4.4          & 0        &    4.4  \\
Lepton ID    & 0     &    1 &    1 &    1 &    1           &    1          & 0        &    1  \\
Lepton Energy Scale    & 0     &    1.5 &    1.5 &    1.5 &    1.5           &    1.5          & 0        &    1.5  \\
$ZH$ Cross Section    & 0     &    0 &    0 &    0 &    0           &    0          & 0        &    5 \\
Fake Leptons       & 50    & 0    & 0    & 0    & 0              & 0             & 0        & 0     \\
Jet Energy Scale  (shape dep.)       & 0     & 
  $^{+0.01}_{-0.01}$   & 
  $^{+0.0}_{-0.0}$   & 
  $^{+0.0}_{-3.2}$   & 
  $^{+5.8}_{-6.3}$   & 
  $^{+7.1}_{-5.8}$   & 
  0   & 
  $^{+2.3}_{+0.0}$   \\ 
Mistag Rate (shape dep.)      & 0     & 0    & 0    & 0    & 0              & 0             &   $^{+31.5}_{-27.2}$     & 0     \\
B-Tag Efficiency      & 0     &    8 &    8 &    8 &    8           &   8          & 0        &    8  \\
$t{\bar{t}}$ Cross Section         & 0     &   20 & 0    & 0    & 0              & 0             & 0        & 0     \\
Diboson Cross Section        & 0     & 0    & 11.5   & 11.5    & 0              & 0             & 0        & 0     \\
$\sigma(p{\bar{p}}\rightarrow Z+HF)$      & 0     & 0    & 0    & 0    &  40            & 40           & 0        & 0     \\
ISR (shape dep.)           & 0     & 0    & 0    & 0    & 0              & 0             & 0        &   $^{-0.01}_{+0.0}$     \\
FSR (shape dep.)           & 0     & 0    & 0    & 0    & 0              & 0             & 0        &   $^{-4.3}_{-0.01}$     \\
\hline
\end{tabular}
\vskip 0.8cm  

{\centerline{CDF: Loose Double Tag High $s/b$ (LDT High)~ $ZH \rightarrow \ell\ell b \bar{b}$ analysis relative uncertainties (\%)}}
\vskip 0.099cm                                                                                                          
\begin{tabular}{|l|c|c|c|c|c|c|c|c|} \hline
Contribution   & ~Fakes~ & ~~~Top~~~  & ~~$WZ$~~ & ~~$ZZ$~~ & ~$Z+b{\bar{b}}$~ & ~$Z+c{\bar{c}}$~& ~$Z+$mistag~ & ~~~$ZH$~~~ \\ \hline
Luminosity ($\sigma_{\mathrm{inel}}(p{\bar{p}})$)          & 0     &    3.8 &    3.8 &    3.8 &    3.8           &    3.8          & 0        &    3.8  \\
Luminosity Monitor        & 0     &    4.4 &    4.4 &    4.4 &    4.4           &    4.4          & 0        &    4.4  \\
Lepton ID    & 0     &    1 &    1 &    1 &    1           &    1          & 0        &    1  \\
Lepton Energy Scale    & 0     &    1.5 &    1.5 &    1.5 &    1.5           &    1.5          & 0        &    1.5  \\
$ZH$ Cross Section    & 0     &    0 &    0 &    0 &    0           &    0          & 0        &    5 \\
Fake Leptons       & 50    & 0    & 0    & 0    & 0              & 0             & 0        & 0     \\
Jet Energy Scale  (shape dep.)       & 0     & 
  $^{+1.3}_{-0.01}$   & 
  $^{+3.1}_{-4.3}$   & 
  $^{+3.1}_{-3.0}$   & 
  $^{+7.5}_{-7.3}$   & 
  $^{+6.2}_{-6.0}$   & 
  0   & 
  $^{+1.9}_{+0.0}$   \\ 
Mistag Rate (shape dep.)      & 0     & 0    & 0    & 0    & 0              & 0             &   $^{+33.6}_{-26.4}$     & 0     \\
B-Tag Efficiency      & 0     &    11 &    11 &    11 &    11           &   11          & 0        &    11  \\
$t{\bar{t}}$ Cross Section         & 0     &   20 & 0    & 0    & 0              & 0             & 0        & 0     \\
Diboson Cross Section        & 0     & 0    & 11.5   & 11.5    & 0              & 0             & 0        & 0     \\
$\sigma(p{\bar{p}}\rightarrow Z+HF)$      & 0     & 0    & 0    & 0    &  40            & 40           & 0        & 0     \\
ISR (shape dep.)           & 0     & 0    & 0    & 0    & 0              & 0             & 0        &   $^{+3.0}_{+0.0}$     \\
FSR (shape dep.)           & 0     & 0    & 0    & 0    & 0              & 0             & 0        &   $^{+1.4}_{-0.0}$     \\
\hline
\end{tabular}
\vskip 0.8cm 
{\centerline{CDF: Loose Double Tag Low $s/b$ (LDT Low)~ $ZH \rightarrow \ell\ell b \bar{b}$ analysis relative uncertainties (\%)}}
\vskip 0.099cm                                                                                                          
\begin{tabular}{|l|c|c|c|c|c|c|c|c|} \hline
Contribution   & ~Fakes~ & ~~~Top~~~  & ~~$WZ$~~ & ~~$ZZ$~~ & ~$Z+b{\bar{b}}$~ & ~$Z+c{\bar{c}}$~& ~$Z+$mistag~ & ~~~$ZH$~~~ \\ \hline
Luminosity ($\sigma_{\mathrm{inel}}(p{\bar{p}})$)          & 0     &    3.8 &    3.8 &    3.8 &    3.8           &    3.8          & 0        &    3.8  \\
Luminosity Monitor        & 0     &    4.4 &    4.4 &    4.4 &    4.4           &    4.4          & 0        &    4.4  \\
Lepton ID    & 0     &    1 &    1 &    1 &    1           &    1          & 0        &    1  \\
Lepton Energy Scale    & 0     &    1.5 &    1.5 &    1.5 &    1.5           &    1.5          & 0        &    1.5  \\
$ZH$ Cross Section    & 0     &    0 &    0 &    0 &    0           &    0          & 0        &    5 \\
Fake Leptons       & 50    & 0    & 0    & 0    & 0              & 0             & 0        & 0     \\
Jet Energy Scale  (shape dep.)       & 0     & 
  $^{+1.7}_{-0.0}$   & 
  $^{-0.0}_{-5.9}$   & 
  $^{+2.9}_{-0.01}$   & 
  $^{+8.2}_{-8.8}$   & 
  $^{+8.1}_{-8.8}$   & 
  0   & 
  $^{+2.7}_{-0.0}$   \\ 
Mistag Rate (shape dep.)      & 0     & 0    & 0    & 0    & 0              & 0             &   $^{+34.5}_{-27.8}$     & 0     \\
B-Tag Efficiency      & 0     &    11 &    11 &    11 &    11           &   11          & 0        &    11  \\
$t{\bar{t}}$ Cross Section         & 0     &   20 & 0    & 0    & 0              & 0             & 0        & 0     \\
Diboson Cross Section        & 0     & 0    & 11.5   & 11.5    & 0              & 0             & 0        & 0     \\
$\sigma(p{\bar{p}}\rightarrow Z+HF)$      & 0     & 0    & 0    & 0    &  40            & 40           & 0        & 0     \\
ISR (shape dep.)           & 0     & 0    & 0    & 0    & 0              & 0             & 0        &   $^{+4.1}_{+7.8}$     \\
FSR (shape dep.)           & 0     & 0    & 0    & 0    & 0              & 0             & 0        &   $^{+23.5}_{+9.9}$     \\
\hline
\end{tabular}
\vskip 0.8cm 
\end{center}
\end{table}


\begin{table}
\caption{Systematic uncertainties on the contributions for D0's $ZH\rightarrow \ell^+\ell^-b{\bar{b}}$ single-tag (ST) channel.
Systematic uncertainties are listed by name; see the original references for a detailed explanation of their meaning and on how they are derived.  
Systematic uncertainties for $ZH$  shown in this table are obtained for $m_H=115$ GeV/$c^2$.
Uncertainties are relative, in percent, and are symmetric unless otherwise indicated. }
\label{tab:d0llbb1}
\vskip 0.8cm                                                                                                          
{\centerline{D0: Single Tag (ST)~ $ZH \rightarrow \ell\ell b \bar{b}$ analysis relative uncertainties (\%)}}
\vskip 0.099cm                                                                                                          
\begin{ruledtabular}
\begin{tabular}{  l  c  c  c  c  c  c  c }                                                                               \\
Contribution & ~~WZ/ZZ~~ &~~Zbb/Zcc~~&~~~Zjj~~~ &~~~~$t\bar{t}$~~~~&  Multijet & ~~~ZH~~~\\ \hline                               
Luminosity                             &  6    &  6    &  6    &  6    &  0    &  6    \\ 
EM ID/Reco eff.                        &  2    &  2    &  2    &  2    &  0    &  2    \\                                      
Muon ID/Reco eff.                      &  2    &  2    &  2    &  2    &  0    &  2    \\                                      
Jet ID/Reco eff.                       &  2    &  2    &  2    &  2    &  0    &  2    \\ 
Jet Energy Scale (shape dep.)          &  5    &  5    &  5    &  5    &  0    &  5    \\                                      
$b$-tagging/taggability                  &  5    &  5    &  5    &  5    &  0    &  5    \\ 
Cross Section                          &  6    & 30    &  6    & 10    &  0    &  6    \\                                      
MC modeling                        &  0    &  4    &  4    &  0    &  0    &  0    \\ 
Instrumental-ZH                        &  0    &  0    &  0    &  0    & 50    &  0    \\ 
\end{tabular}                                                                                                           
\end{ruledtabular} 
\vskip 0.8cm                                                                                                          
{\centerline{D0: Double Tag (DT)~ $ZH \rightarrow \ell\ell b \bar{b}$ analysis relative uncertainties (\%)}}
\vskip 0.099cm                                                                                                          
\begin{ruledtabular}
\begin{tabular}{  l  c  c  c  c  c  c  c }  \\                                                                             

Contribution & ~~WZ/ZZ~~ &~~Zbb/Zcc~~&~~~Zjj~~~ &~~~~$t\bar{t}$~~~~&  Multijet & ~~~ZH~\\ \hline                               
Luminosity                             &  6    &  6    &  6    &  6    &  0    &  6    \\ 
EM ID/Reco eff.                        &  2    &  2    &  2    &  2    &  0    &  2    \\                                      
Muon ID/Reco eff.                      &  2    &  2    &  2    &  2    &  0    &  2    \\                                      
Jet ID/Reco eff.                       &  2    &  2    &  2    &  2    &  0    &  2    \\ 
Jet Energy Scale (shape dep.)          &  5    &  5    &  5    &  5    &  0    &  5    \\                                      
$b$-tagging/taggability                  & 10    & 10    & 10    & 10    &  0    & 10    \\ 
Cross Section                          &  6    & 30    &  6    & 10    &  0    &  6    \\                                      
MC modeling                        &  0    &  4    &  4    &  0    &  0    &  0    \\ 
Instrumental-ZH                        &  0    &  0    &  0    &  0    & 50    &  0    \\
\end{tabular}                                                                                                           
\end{ruledtabular}                                                                                                           
\end{table}

\clearpage

\begin{table}
\caption{Systematic uncertainties on the contributions for CDF's $H\rightarrow W^+W^-\rightarrow\ell^{\pm}\ell^{\prime \mp}$ channels with zero, 
one, and two or more associated jets.  These channels are sensitive to gluon fusion production (all channels) and $WH, ZH$ and VBF production 
(channels with one or more associated jets).  Systematic uncertainties are listed by name (see the original references for a detailed explanation 
of their meaning and on how they are derived).  Systematic uncertainties for $H$ shown in this table are obtained for $m_H=160$ GeV/$c^2$.
Uncertainties are relative, in percent, and are symmetric unless otherwise indicated.  The uncertainties associated with the different background 
and signal processed are correlated within individual jet categories unless otherwise noted.  Boldface and italics indicate groups of uncertainties
which are correlated with each other but not the others on the line. 
Monte Carlo statistical uncertainties in each bin of each template are considered as independent 
systematic uncertainties.  All uncertainty categories are treated as correlated between channels with the exception of the Missing Et Modeling 
uncertainty.}
\label{tab:cdfsystww0}
\vskip 0.3cm                                                                             
{\centerline{CDF: $H\rightarrow W^+W^-\rightarrow\ell^{\pm}\ell^{\prime \mp}$ channels with no associated jet relative uncertainties (\%)}}
\vskip 0.099cm
\begin{ruledtabular}
\begin{tabular}{lcccccccc}\\
Uncertainty Source      & $WW$       & $WZ$       & $ZZ$       & $t\bar{t}$ & DY      & $W\gamma$  & $W$+jet(s) & $gg\to H$ \\ \hline 
{Cross Section}         & {\it 6.0}  & {\it 6.0}  & {\it 6.0}  & 10.0       & 5.0     &            &            & 10.4      \\ \hline
Scale (leptons)         &            &            &            &            &         &            &            & 2.5       \\ 
Scale (jets)            &            &            &            &            &         &            &            & 4.6       \\ 
PDF Model (leptons)     & 1.9        & 2.7        & 2.7        & 2.1        & 4.1     &            &            & 1.5       \\ 
PDF Model (jets)        &            &            &            &            &         &            &            & 0.9       \\ 
Higher-order Diagrams   & {\it 5.0}  & {\it 10.0} & {\it 10.0} & 10.0       &         & 11.0       &            &           \\ 
Missing Et Modeling     &            &            &            &            & 21.0    &            &            &           \\ 
$W\gamma$ Scaling       &            &            &            &            &         & 12.0       &            &           \\ 
Jet Fake Rates   
(Low/High $s/b$)        &            &            &            &            &         &            & 21.5/27.7  &           \\
Jet Modeling            & -1.0       &            &            &            &         & {\it -4.0} &            &           \\
MC Run Dependence       & 2.8        &            &            &            &         &            &            &           \\ 
Lepton ID Efficiencies  & 2.0        & 1.7        & 2.0        & 2.0        & 1.9     &            &            & 1.9       \\ 
Trigger Efficiencies    & 2.1        & 2.1        & 2.1        & 2.0        & 3.4     &            &            & 3.3       \\ \hline
Luminosity              & 3.8        & 3.8        & 3.8        & 3.8        & 3.8     &            &            & 3.8       \\ 
Luminosity Monitor      & 4.4        & 4.4        & 4.4        & 4.4        & 4.4     &            &            & 4.4       \\ 
\end{tabular}
\end{ruledtabular}
\end{table}

\begin{table}
{\centerline{CDF: $H\rightarrow W^+W^-\rightarrow\ell^{\pm}\ell^{\prime \mp}$ channels with one associated jet relative uncertainties (\%)}}
\vskip 0.099cm
\begin{ruledtabular}
\begin{tabular}{lccccccccccc} \\
Uncertainty Source      & $WW$       & $WZ$       & $ZZ$       & $t\bar{t}$ & DY      & $W\gamma$  & $W$+jet(s) & $gg\to H$ & $WH$       & $ZH$       & VBF        \\ \hline 
Cross Section           & {\it 6.0}  & {\it 6.0}  & {\it 6.0}  & 10.0       & 5.0     &            &            & 24.7      & {\bf 5.0}  & {\bf 5.0}  & 10.0       \\ \hline
Scale (leptons)         &            &            &            &            &         &            &            & 2.8       &            &            &            \\ 
Scale (jets)            &            &            &            &            &         &            &            & -5.1      &            &            &            \\ 
PDF Model (leptons)     & 1.9        & 2.7        & 2.7        & 2.1        & 4.1     &            &            & 1.7       & 1.2        & 0.9        & 2.2        \\ 
PDF Model (jets)        &            &            &            &            &         &            &            & -1.9      &            &            &            \\ 
Higher-order Diagrams   & {\it 5.0}  & {\it 10.0} & {\it 10.0} & 10.0       &         & 11.0       &            &           & {\bf 10.0} & {\bf 10.0} & {\bf 10.0} \\ 
Missing Et Modeling     &            &            &            &            & 30.0    &            &            &           &            &            &            \\ 
$W\gamma$ Scaling       &            &            &            &            &         & 12.0       &            &           &            &            &            \\ 
Jet Fake Rates 
(Low/High $s/b$)        &            &            &            &            &         &            & 22.2/31.5  &           &            &            &            \\
Jet Modeling            & -1.0       &            &            &            &         & {\it 15.0} &            &           &            &            &            \\
MC Run Dependence       & 1.0        &            &            &            &         &            &            &           &            &            &            \\ 
Lepton ID Efficiencies  & 2.0        & 2.0        & 2.2        & 1.8        & 2.0     &            &            & 1.9       & 1.9        & 1.9        & 1.9        \\ 
Trigger Efficiencies    & 2.1        & 2.1        & 2.1        & 2.0        & 3.4     &            &            & 3.3       & 2.1        & 2.1        & 3.3        \\ \hline
Luminosity              & 3.8        & 3.8        & 3.8        & 3.8        & 3.8     &            &            & 3.8       & 3.8        & 3.8        & 3.8        \\
Luminosity Monitor      & 4.4        & 4.4        & 4.4        & 4.4        & 4.4     &            &            & 4.4       & 4.4        & 4.4        & 4.4        \\ 
\end{tabular}
\end{ruledtabular}
\end{table}

\begin{table}
{\centerline{CDF: $H\rightarrow W^+W^-\rightarrow\ell^{\pm}\ell^{\prime \mp}$ channel with two or more associated jets relative uncertainties (\%)}}
\vskip 0.0999cm
\label{tab:cdfsystww2}
\begin{ruledtabular}
\begin{tabular}{lccccccccccc}\\
Uncertainty Source      & $WW$       & $WZ$       & $ZZ$       & $t\bar{t}$ & DY      & $W\gamma$  & $W$+jet(s) & $gg\to H$ & $WH$       & $ZH$       & VBF        \\ \hline 
Cross Section           & {\it 6.0}  & {\it 6.0}  & {\it 6.0}  & 10.0       & 5.0     &            &            & 67.9      & {\bf 5.0}  & {\bf 5.0}  & 10.0       \\ \hline
Scale (leptons)         &            &            &            &            &         &            &            & 3.1       &            &            &            \\ 
Scale (jets)            &            &            &            &            &         &            &            & -8.7      &            &            &            \\ 
PDF Model (leptons)     & 1.9        & 2.7        & 2.7        & 2.1        & 4.1     &            &            & 2.0       & 1.2        & 0.9        & 2.2        \\ 
PDF Model (jets)        &            &            &            &            &         &            &            & -2.8      &            &            &            \\ 
Higher-order Diagrams   & {\it 5.0}  & {\it 10.0} & {\it 10.0} & 10.0       &         & 11.0       &            &           & {\bf 10.0} & {\bf 10.0} & {\bf 10.0} \\ 
Missing Et Modeling     &            &            &            &            & 32.0    &            &            &           &            &            &            \\ 
$W\gamma$ Scaling       &            &            &            &            &         & 12.0       &            &           &            &            &            \\ 
Jet Fake Rates          &            &            &            &            &         &            & 27.1       &           &            &            &            \\
Jet Modeling            & 20.0       &            &            &            &         & {\it 18.5} &            &           &            &            &            \\
$b$-tag veto            &            &            &            & 5.4        &         &            &            &           &            &            &            \\
MC Run Dependence       & 1.5        &            &            &            &         &            &            &           &            &            &            \\ 
Lepton ID Efficiencies  & 1.9        & 2.9        & 1.9        & 1.9        & 1.9     &            &            & 1.9       & 1.9        & 1.9        & 1.9        \\ 
Trigger Efficiencies    & 2.1        & 2.1        & 2.1        & 2.0        & 3.4     &            &            & 3.3       & 2.1        & 2.1        & 3.3        \\ \hline
Luminosity              & 3.8        & 3.8        & 3.8        & 3.8        & 3.8     &            &            & 3.8       & 3.8        & 3.8        & 3.8        \\
Luminosity Monitor      & 4.4        & 4.4        & 4.4        & 4.4        & 4.4     &            &            & 4.4       & 4.4        & 4.4        & 4.4        \\ 
\end{tabular}                                                                                                             
\end{ruledtabular}                                                                                                             
\end{table}

\begin{table}
\caption{Systematic uncertainties on the contributions for CDF's $H\rightarrow W^+W^-\rightarrow\ell^{\pm}\ell^{\prime \mp}$ low-$M_{\ell\ell}$ channel with zero
or one associated jets.  This channel is sensitive to only gluon fusion production.  Systematic uncertainties are listed by name (see the original references for 
a detailed explanation of their meaning and on how they are derived).  Systematic uncertainties for $H$ shown in this table are obtained for $m_H=160$ GeV/$c^2$.
Uncertainties are relative, in percent, and are symmetric unless otherwise indicated.  The uncertainties associated with the different background and signal 
processed are correlated within individual categories unless otherwise noted.  In these special cases, the correlated uncertainties are shown in either italics 
or bold face text.  Monte Carlo statistical uncertainties in each bin of each template are considered as independent systematic uncertainties.  All uncertainty 
categories are treated as correlated between channels with the exception of the Missing Et Modeling uncertainty.}
\label{tab:cdfsystww1}
\vskip 0.3cm                                                                             
{\centerline{CDF: $H\rightarrow W^+W^-\rightarrow\ell^{\pm}\ell^{\prime \mp}$ low $M_{\ell\ell}$ channel with zero or one associated jets relative uncertainties (\%)}}
\vskip 0.099cm
\begin{ruledtabular}
\begin{tabular}{lcccccccc}\\
Uncertainty Source      & $WW$       & $WZ$       & $ZZ$       & $t\bar{t}$ & DY      & $W\gamma$  & $W$+jet(s) & $gg\to H$ \\ \hline 
{Cross Section}         & {\it 6.0}  & {\it 6.0}  & {\it 6.0}  & 10.0       & 5.0     &            &            & 14.3      \\ \hline
Scale (leptons)         &            &            &            &            &         &            &            & 2.6       \\ 
Scale (jets)            &            &            &            &            &         &            &            & 1.1       \\ 
PDF Model (leptons)     & 1.9        & 2.7        & 2.7        & 2.1        & 4.1     &            &            & 1.7       \\ 
PDF Model (jets)        &            &            &            &            &         &            &            & 0.3       \\ 
Higher-order Diagrams   & {\it 5.5}  & {\it 11.0} & {\it 11.0} & 10.0       &         &            &            &           \\ 
Missing Et Modeling     &            &            &            &            & 22.0    &            &            &           \\ 
$W\gamma$ Scaling       &            &            &            &            &         & 12.0       &            &           \\ 
Jet Fake Rates          &            &            &            &            &         &            & 24.1       &           \\
Jet Modeling            & -1.0       &            &            &            &         &            &            &           \\
MC Run Dependence       & 5.0        &            &            &            &         &            &            &           \\ 
Lepton ID Efficiencies  & 2.0        & 1.7        & 2.0        & 2.0        & 1.9     &            &            & 1.9       \\ 
Trigger Efficiencies    & 2.1        & 2.1        & 2.1        & 2.0        & 3.4     &            &            & 3.3       \\ \hline
Luminosity              & 3.8        & 3.8        & 3.8        & 3.8        & 3.8     &            &            & 3.8       \\ 
Luminosity Monitor      & 4.4        & 4.4        & 4.4        & 4.4        & 4.4     &            &            & 4.4       \\ 
\end{tabular}
\end{ruledtabular}
\end{table}
%


\begin{table}
\caption{
Systematic uncertainties on the contributions for CDF's $WH\rightarrow WWW \rightarrow\ell^{\prime \pm}\ell^{\prime \pm}$ channel with 
one or more associated jets.  This channel is sensitive to only $WH$ and $ZH$ production.  Systematic uncertainties are listed by name 
(see the original references for a detailed explanation of their meaning and on how they are derived).  Systematic uncertainties for 
$H$ shown in this table are obtained for $m_H=160$ GeV/$c^2$.  Uncertainties are relative, in percent, and are symmetric unless otherwise 
indicated.  The uncertainties associated with the different background and signal processed are correlated within individual categories 
unless otherwise noted.  In these special cases, the correlated uncertainties are shown in either italics or bold face text.  Monte Carlo 
statistical uncertainties in each bin of each template are considered as independent systematic uncertainties.  All uncertainty categories 
are treated as correlated between channels with the exception of the Missing Et Modeling uncertainty.}
\vglue 0.2cm
\label{tab:cdfsystwww}
\vskip 0.8cm                                                                                                          
{\centerline{CDF: $WH \rightarrow WWW \rightarrow\ell^{\pm}\ell^{\prime\pm}$ analysis relative uncertainties (\%)}}
\vskip 0.099cm                                                                                                          
\begin{ruledtabular}
\begin{tabular}{lccccccccc}\\
Uncertainty Source      & $WW$       & $WZ$       & $ZZ$       & $t\bar{t}$ & DY      & $W\gamma$  & $W$+jet(s) & $WH$       & $ZH$       \\ \hline 
Cross Section           & {\it 6.0}  & {\it 6.0}  & {\it 6.0}  & 10.0       & 5.0     &            &            & {\bf 5.0}  & {\bf 5.0}  \\ \hline
Scale (leptons)         &            &            &            &            &         &            &            &            &            \\ 
Scale (jets)            &            &            &            &            &         &            &            &            &            \\ 
PDF Model (leptons)     & 1.9        & 2.7        & 2.7        & 2.1        & 4.1     &            &            & 1.2        & 0.9        \\ 
PDF Model (jets)        &            &            &            &            &         &            &            &            &            \\ 
Higher-order Diagrams   & {\it 5.0}  & {\it 10.0} & {\it 10.0} & 10.0       &         & 11.0       &            & {\bf 10.0} & {\bf 10.0} \\ 
Missing Et Modeling     &            &            &            &            & 17.0    &            &            &            &            \\ 
$W\gamma$ Scaling       &            &            &            &            &         & 12.0       &            &            &            \\ 
Jet Fake Rates          &            &            &            &            &         &            & 30.0       &            &            \\
Jet Modeling            & 3.0        &            &            &            &         & {\it 16.0} &            &            &            \\
Charge Misassignment    & 16.5       &            &            & 16.5       & 16.5    &            &            &            &            \\
MC Run Dependence       & 1.0        &            &            &            &         &            &            &            &            \\ 
Lepton ID Efficiencies  & 2.0        & 2.0        & 2.0        & 2.0        & 2.0     &            &            & 2.0        & 2.0        \\ 
Trigger Efficiencies    & 2.1        & 2.1        & 2.1        & 2.0        & 3.4     &            &            & 2.1        & 2.1        \\ \hline
Luminosity              & 3.8        & 3.8        & 3.8        & 3.8        & 3.8     &            &            & 3.8        & 3.8        \\
Luminosity Monitor      & 4.4        & 4.4        & 4.4        & 4.4        & 4.4     &            &            & 4.4        & 4.4        \\ 
\end{tabular}                                                                                                             
\end{ruledtabular}                                                                                                             
\end{table}

\begin{table}
\caption{Systematic uncertainties on the contributions for D0's
$WH \rightarrow WWW \rightarrow\ell^{\prime \pm}\ell^{\prime \pm}$ channel.
Systematic uncertainties are listed by name; see the original references for a detailed explanation of their meaning and on how they are derived. 
Systematic uncertainties for $WH$ shown in this table are obtained for $m_H=165$ GeV/$c^2$.
Uncertainties are relative, in percent, and are symmetric unless otherwise indicated.   }
\label{tab:d0systwww}
\vskip 0.8cm                                                                                                          
{\centerline{D0: $WH \rightarrow WWW \rightarrow\ell^{\pm}\ell^{\prime\pm}$ Run IIa analysis relative uncertainties (\%)}}
\vskip 0.099cm                                                                                                   
\begin{ruledtabular}       
\begin{tabular}{ l  c  c  c  c  }\\

Contribution                           & ~~WZ/ZZ~~ & Charge flips & ~Multijet~ &~~~~~WH~~~~~  \\ \hline
Luminosity                             & 6 &  0 &  0   & 6 \\
Trigger eff.                           &  5    &  0                     &  0  &  5    \\
Lepton ID/Reco. eff                    & 10    &  0                     &  0  & 10    \\
Cross Section                          &  7    &  0                     &  0  &  6    \\
Normalization                          &  6    &  0                     &  0  &  0    \\ 
Instrumental-$ee$ ($ee$ final state)                                                    
                                       &  0    &  32                    &  15 &  0    \\
Instrumental-$e\mu$ ($e\mu$ final state)                                                  
                                       &  0    &  0                     &  18 &  0    \\
Instrumental-$\mu\mu$ ($\mu\mu$ final state)                                                 
                                       &  0    &  $^{+290}_{-100}$      & 32  &  0    \\ \hline

\end{tabular}
\end{ruledtabular}
\vskip 0.5cm
{\centerline{D0: $WH \rightarrow WWW
\rightarrow\ell^{\pm}\ell^{\prime\pm}$ Run IIb analysis relative uncertainties (\%)}}
\vskip 0.099cm
\begin{ruledtabular}
\begin{tabular}{ l  c  c  c  c  }
\hline
Contribution                & WZ/ZZ     & Charge flips & Multijet     &WH  \\ 
\hline
Luminosity & 6 &  0 &  0   & 6 \\
Lepton ID -$ee$            &  9        &  0                &  0      & 9    \\
Lepton ID -$\mu\mu$       & 4        &  0                 &  0       & 4    \\
Lepton ID -$e\mu$          &  5        &  0                &  0      & 5    \\
Cross Section               &  7        &  0                &  0      &  6    \\
Instrumental-$ee$ ($ee$ final state)          &  0        &  60                &  95    &  0    \\
Instrumental-$\mu\mu$   ($\mu\mu$ final state)  &  0        &  100               &  140    &  0    \\
Instrumental-$e\mu$   ($e\mu$ final state)     &  0         &  0                 & 9      &  0    \\
\hline
\end{tabular}
\end{ruledtabular}
\end{table}

\clearpage

\begin{table}
\caption{Systematic uncertainties on the contributions for D0's
$H\rightarrow WW \rightarrow\ell^{\pm}\ell^{\prime \mp}$ channels.
Systematic uncertainties are listed by name; see the original references for a detailed explanation of their meaning and on how they are derived.
Systematic uncertainties shown in this table are obtained for the $m_H=165$ GeV/c$^2$ Higgs selection.
Uncertainties are relative, in percent, and are symmetric unless otherwise indicated.   }
\label{tab:d0systww}
\vskip 0.8cm
{\centerline{D0: $H\rightarrow WW \rightarrow e^{\pm} e^{ \mp}$ analysis  relative uncertainties (\%)}}
\vskip 0.099cm
\begin{ruledtabular}
\begin{tabular}{ l  c  c  c  c  c  c  c}  \\

Contribution & Diboson & ~~$Z/\gamma^* \rightarrow \ell\ell$~~&$~~W+jet/\gamma$~~ &~~~~$t\bar{t}~~~~$    & ~~Multijet~~  & ~~~~$H$~~~~      \\
\hline
Lepton ID                        &  6           &   6           & 6             & 6            & --   &   6        \\
Charge mis-ID                    &  1           &   1           & 1             & 1            & --   &   1        \\
Jet Energy Scale (s)               &  1           &   1        & 1          & 1          & --   &   1      \\
Jet identification (s)              &  1           &   1           & 1             & 1            & --   &   1        \\
Cross Section                     &  7           &   7           & 7            & 10           & 2  &   11         \\
Luminosity                       &  6           &   6           & 6             & 6            & --   &   6      \\
Modeling (s)~~~~~                  &  0           &   1           & 1             & 0            &  --   &   1     \\

\end{tabular}
\end{ruledtabular}
\vskip 0.8cm
{\centerline{D0: $H\rightarrow WW \rightarrow e^{\pm} \mu^{\mp}$ analysis  relative uncertainties (\%)}}
\vskip 0.099cm
\begin{ruledtabular}
\begin{tabular}{ l  c  c  c  c  c  c  c} \\

Contribution & Diboson & ~~$Z/\gamma^* \rightarrow \ell\ell$~~&$~~W+jet/\gamma$~~ &~~~~$t\bar{t}~~~~$    & ~~Multijet~~  & ~~~~$H$~~~~      \\
\hline
Trigger                          &  2           &   2           & 2             & 2            & --   &   2         \\
Lepton ID                        &  3           &   3           & 3             & 3            & --   &   3        \\
Momentum resolution (s)          &  0           &   3           & 1             & 0            & --   &   0       \\
Jet Energy Scale (s)              &  1           &   5           & 1             & 1            & --   &   1       \\
Jet identification (s)              &  1            &   3           & 1             & 1            & --   &   1        \\
Cross Section                     &  7           &   7           & 7            & 10           & 10   &   11       \\
Luminosity                       &  6           &   6           & 6             & 6            & --   &   6      \\
Modeling (s)~~~~~                  &  1           &   1           & 3             & 0            &  0   &   1      \\

\end{tabular}
\end{ruledtabular}
\label{tab:d0systww_mm}
\vskip 0.8cm
{\centerline{D0: $H\rightarrow WW \rightarrow \mu^{\pm} \mu^{\mp}$ analysis  relative uncertainties (\%)}}
\vskip 0.099cm
\begin{ruledtabular}
\begin{tabular}{ l  c  c  c  c  c  c  c} \\

Contribution & Diboson & ~~$Z/\gamma^* \rightarrow \ell\ell$~~&$~~W+jet/\gamma$~~ &~~~~$t\bar{t}~~~~$    & ~~Multijet~~  & ~~~~$H$~~~~      \\
\hline
Lepton ID                        &  4           &   4           & 4             & 4            & --   &   4      \\
Momentum resolution (s)          &  1           &   1           & 2             & 1            & --   &   1      \\
Charge mis-ID                    &  1           &   1           & 1             & 1            & --   &   1      \\
Jet Energy Scale (s)             &  1           &   1           & 1             & 1            & --   &   1     \\
Jet identification               &  1           &   1           & 3             & 1            & --   &   1      \\
Cross Section                    &  7           &   7           & 7            & 10           &  15   &   11     \\
Luminosity                       &  6           &   6           & 6             & 6            & --   &   6     \\
Modeling~~~~~                    &  0           &   0           & 1             & 0            &  0   &   1      \\

\end{tabular}
\end{ruledtabular}
\end{table}

\begin{table}
\caption{Systematic uncertainties on the contributions for D0's
  $ t \bar{t} H\rightarrow t \bar{t} b \bar{b}$ channel.
Systematic uncertainties for $ZH$, $WH$  shown in this table are obtained for $m_H=115$ GeV/$c^2$.
Systematic uncertainties are listed by name; see the original references for a detailed explanation of 
their meaning and on how they are derived.  %
Uncertainties are relative, in percent, and are symmetric unless otherwise indicated.   
}
\label{tab:d0systtth}
\vskip 0.3cm                                                                                                          
{\centerline{D0:   $ t \bar{t} H\rightarrow t \bar{t} b \bar{b}$  analysis relative uncertainties (\%)}}
\vskip 0.099cm                                                                                                          
\begin{ruledtabular}
\begin{tabular}{lcc}\\
Contribution                             &  ~~~background~~~ & ~~~$t \bar{t} H$~~~    \\  
\hline
Luminosity~~~~                           &  6          &  6    \\ 
lepton ID efficiency                     &  2--3       &  2--3    \\
Event preselection                       &  1          &  1    \\
$W$ +jet modeling                        &   15        & -     \\
Cross Section                            &  10--50     &  10    \\
\end{tabular}
\end{ruledtabular}
\end{table}

\begin{table}
\caption{Systematic uncertainties on the contributions for CDF's
$ H\rightarrow \tau^+\tau^-$ channels.
Systematic uncertainties are listed by name; see the original references for a detailed explanation of 
their meaning and on how they are derived.  %
Systematic uncertainties for $H$ shown in this table are obtained for $m_H=115$ GeV/$c^2$.
Uncertainties are relative, in percent, and are symmetric unless otherwise indicated.   The systematic uncertainty 
called ``Normalization''  includes effects of the inelastic $p{\bar{p}}$ cross section, the luminosity monitor acceptance, and the lepton trigger
acceptance. It is considered to be entirely correlated with the luminosity uncertainty.
}
\label{tab:cdfsysttautau}
\vskip 0.3cm                                                                                                          
{\centerline{CDF:   $H \rightarrow \tau^+ \tau^-$ analysis relative uncertainties (\%)}}
\vskip 0.099cm                                                                                                          
\begin{ruledtabular}
\begin{tabular}{lcccccccccc}\\
Contribution & $Z/\gamma^* \rightarrow \tau\tau$ & $Z/\gamma^* \rightarrow \ell\ell$ & $t\bar{t}$ & diboson  & jet $\rightarrow \tau$ 
& W+jet & $WH$      & $ZH$  & VBF      & $gg\rightarrow H$    \\  
\hline
Luminosity                    &  3.8   &  3.8   &  3.8   &  3.8  &  -   &    -     &  3.8  &  3.8  &  3.8  &  3.8   \\ 
Luminosity Monitor            &  4.4   &  4.4   &  4.4   &  4.4  &  -   &    -     &  4.4  &  4.4  &  4.4  &  4.4   \\ 
$e,\mu$ Trigger               &  1   &  1   &  1   &  1  &  -   &    -     &  1  &  1  &  1  &  1   \\
$\tau$  Trigger               &  3   &  3   &  3   &  3  &  -   &    -     &  3  &  3  &  3  &  3   \\
$e,\mu,\tau$ ID               &  3   &  3   &  3   &  3  &  -   &    -     &  3  &  3  &  3  &  3   \\
PDF Uncertainty               &  1   &  1   &  1   &  1  &  -   &    -     &  1  &  1  &  1  &  1   \\
ISR/FSR                       &  -   &  -   &  -   &  -  &  -   &    -     & 2/0 & 1/1 & 3/1 & 12/1 \\
JES (shape)                   & 16   & 13   &  2   & 10  &  -   &    -     &  3  &  3  &  4  & 14   \\
Cross Section or Norm.        &  2  &  2   & 10   & 11.5  &  -   &   15     &  5  &  5  & 10  & 67.9   \\
MC model                      & 20   & 10   &  -   &  -  &  -   &    -     &  -  &  -  &  -  &  -   \\
\end{tabular}
\end{ruledtabular}
\end{table}

\begin{table}
\label{tab:cdfjjbbsyst}
\caption[]{Systematic uncertainties summary for CDF's $WH+ZH\rightarrow jjbb$ channel.
Systematic uncertainties are listed by name; see the original references for a detailed explanation of 
their meaning and on how they are derived. 
Uncertainties with provided shape systematics are labeled with ``s''.
Systematic uncertainties for $H$ shown in this table are obtained for $m_H=115$ GeV/$c^2$.
Uncertainties are relative, in percent, and are symmetric unless otherwise indicated.   
The cross section uncertainties are uncorrelated with each other (except for single top and $t{\bar{t}}$, which are
treated as correlated).  The QCD uncertainty is also uncorrelated with other channels' QCD rate uncertainties.
}
\vskip 0.3cm                                                                                                          
{\centerline{CDF: $WH+ZH\rightarrow jjbb$ analysis relative uncertainties (\%)}}
\vskip 0.099cm
\begin{ruledtabular}
\begin{tabular}{lccccccccc}\\
               &  QCD   &  $t{\bar{t}}$   &  $Wb{\bar{b}}$   &  $WZ$   &  Single Top &  $Z$+jets   &  $WH$   &  $ZH$   \\\hline
 Interpolation & 0s     & --              & --               & --      &  --         &  --         &   --    &   -- \\
 MC Modeling   & 0s     & --              & --               & --      &  --         &  --         &   18s   &  16s \\
 Cross Section & 10     & 10              & 30               & 6       &  10         & 30          & 5       &  5 \\
\hline
\end{tabular}
\end{ruledtabular}
\end{table}

\begin{table}
\caption{Systematic uncertainties on the contributions for D0's
$ H\rightarrow \gamma \gamma$ channels.
Systematic uncertainties for $ZH$, $WH$  shown in this table are obtained for $m_H=115$ GeV/$c^2$.
Systematic uncertainties are listed by name; see the original references for a detailed explanation of 
their meaning and on how they are derived.  %
Uncertainties are relative, in percent, and are symmetric unless otherwise indicated.   
}
\label{tab:d0systgg}
\vskip 0.3cm                                                                                                          
{\centerline{D0:   $H \rightarrow \gamma \gamma$ analysis relative uncertainties (\%)}}
\vskip 0.099cm       
                                                                                                   
\begin{ruledtabular}
\begin{tabular}{lcc}\\
Contribution &  ~~~background~~~  & ~~~$H$~~~    \\  
\hline
Luminosity~~~~                            &  6     &  6    \\ 
Acceptance                                &  --    &   2   \\
electron ID efficiency                    &  2     &  2    \\
electron track-match inefficiency         & 10--20 & -     \\
Photon ID efficiency                      &  7     &   7     \\
Photon energy scale                       &  --    &   2     \\
Acceptance                                &  --    &  2    \\
$\gamma$-jet and jet-jet fakes          &  26    &  --    \\
Cross Section ($Z$)                       &  4     &  6    \\
Background subtraction                    &  8--14 &  -    \\
\hline
\end{tabular}
\end{ruledtabular}
\end{table}

\clearpage

 \begin{figure}[t]
 \begin{centering}
 \includegraphics[width=14.0cm]{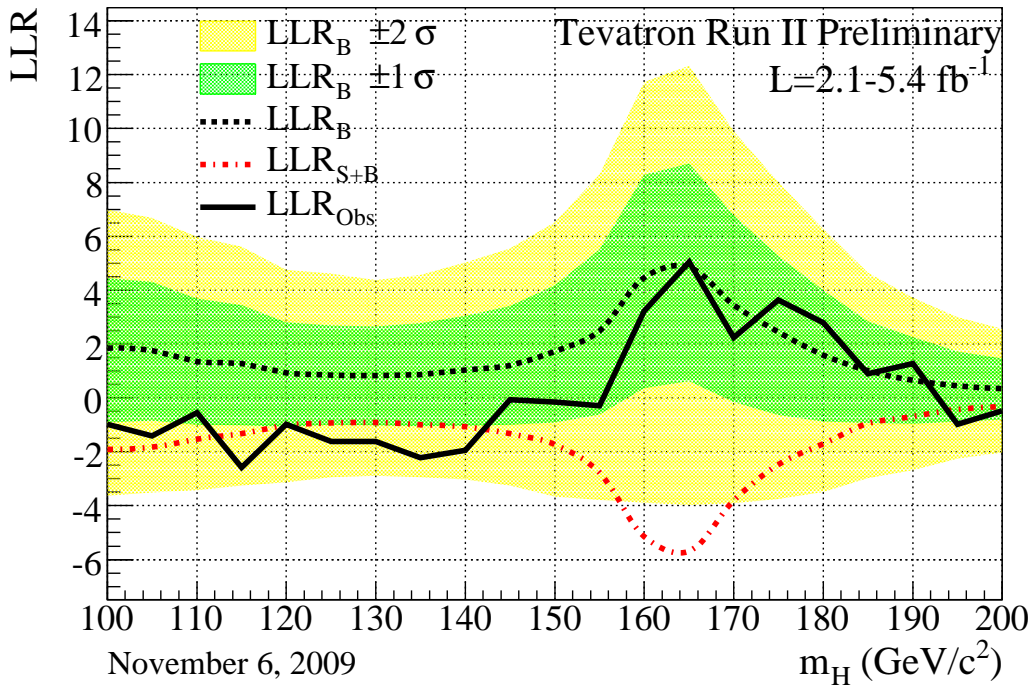}
 \caption{
 \label{fig:comboLLR} { 
Distributions of the log-likelihood ratio (LLR) as a function of Higgs mass obtained with the $CL_s$ method for the combination of all CDF and D0 analyses. 
}}
 \end{centering}
 \end{figure}

\vspace*{1cm}
\section{Combined Results} 

Before extracting the combined limits we study the distributions of the 
log-likelihood ratio (LLR) for different hypotheses, to quantify the expected
sensitivity across the mass range tested.
Figure~\ref{fig:comboLLR} displays the LLR distributions for the combined
analyses as functions of $m_{H}$. Included are the median of the LLR distributions for the
background-only hypothesis (LLR$_{b}$), the signal-plus-background
hypothesis (LLR$_{s+b}$), and the observed value for the data (LLR$_{\rm{obs}}$).  The
shaded bands represent the one and two standard deviation ($\sigma$)
departures for LLR$_{b}$ centered on the median. Table~\ref{tab:llrVals} lists the observed
and expected LLR values shown in Figure~\ref{fig:comboLLR}.


\begin{table}[htpb]
\caption{\label{tab:llrVals} Log-likelihood ratio (LLR) values for the combined CDF + \Dzero Higgs boson search obtained using the CL$_{S}$ method.}
\begin{ruledtabular}
\begin{tabular}{lccccccc} 
$m_{H}$ (GeV/$c^2$ &  LLR$_{\rm{obs}}$ & LLR$_{S+B}^{\rm{med}}$ &  
LLR$_{B}^{-2\sigma}$ & LLR$_{B}^{-1\sigma}$ & LLR$_{B}^{\rm{med}}$ &  LLR$_{B}^{+1\sigma}$ & LLR$_{B}^{+2\sigma}$ \\ \hline
100 & -0.99 & -1.91 & 6.99 & 4.45 & 1.85 & -0.91 & -3.63 \\ 
105 & -1.41 & -1.83 & 6.69 & 4.31 & 1.75 & -0.85 & -3.51 \\ 
110 & -0.55 & -1.53 & 5.97 & 3.69 & 1.35 & -0.99 & -3.43 \\ 
115 & -2.58 & -1.33 & 5.61 & 3.45 & 1.27 & -1.01 & -3.25 \\ 
120 & -0.99 & -1.01 & 4.75 & 2.81 & 0.93 & -1.07 & -3.13 \\ 
125 & -1.62 & -0.93 & 4.61 & 2.69 & 0.85 & -0.99 & -2.95 \\ 
130 & -1.61 & -0.91 & 4.37 & 2.65 & 0.83 & -1.01 & -2.89 \\ 
135 & -2.22 & -0.99 & 4.57 & 2.77 & 0.87 & -1.05 & -2.95 \\ 
140 & -1.94 & -1.07 & 5.03 & 3.05 & 1.03 & -1.03 & -3.03 \\ 
145 & -0.07 & -1.33 & 5.55 & 3.41 & 1.21 & -0.99 & -3.25 \\ 
150 & -0.15 & -1.73 & 6.53 & 4.19 & 1.73 & -0.91 & -3.67 \\ 
155 & -0.29 & -2.77 & 8.31 & 5.51 & 2.49 & -0.61 & -3.79 \\ 
160 & 3.23 & -5.15 & 11.73 & 8.29 & 4.47 & 0.37 & -3.89 \\ 
165 & 5.04 & -5.69 & 12.33 & 8.69 & 4.85 & 0.63 & -3.97 \\ 
170 & 2.24 & -3.81 & 9.91 & 6.79 & 3.45 & -0.15 & -3.91 \\ 
175 & 3.64 & -2.47 & 8.01 & 5.27 & 2.45 & -0.63 & -3.75 \\ 
180 & 2.79 & -1.71 & 6.27 & 3.99 & 1.59 & -0.87 & -3.49 \\ 
185 & 0.90 & -0.95 & 4.65 & 2.85 & 1.01 & -0.93 & -2.97 \\ 
190 & 1.28 & -0.69 & 3.73 & 2.25 & 0.65 & -0.97 & -2.69 \\ 
195 & -0.98 & -0.43 & 2.99 & 1.73 & 0.45 & -0.87 & -2.25 \\ 
200 & -0.48 & -0.33 & 2.57 & 1.47 & 0.35 & -0.81 & -2.03 
\end{tabular}
\end{ruledtabular}
\end{table}

These
distributions can be interpreted as follows:
The separation between the medians of the LLR$_{b}$ and LLR$_{s+b}$ distributions provides a
measure of the discriminating power of the search.  The sizes  
of the one- and two-$\sigma$ LLR$_{b}$ bands indicate the width of the LLR$_{b}$ distribution,
assuming no signal is truly present and only statistical fluctuations and systematic effects are
present.  The value of LLR$_{\rm{obs}}$ relative to LLR$_{s+b}$ and LLR$_{b}$
indicates whether the data distribution appears to resemble what we expect if a signal is present
(i.e. closer to the LLR$_{s+b}$ distribution, which is negative by
construction)
or whether it resembles the background expectation more closely; the significance of any departures
of LLR$_{\rm{obs}}$ from LLR$_{b}$ can be evaluated by the width of the
LLR$_{b}$ bands.

Using the combination procedures outlined in Section III, we extract limits on
SM Higgs boson production $\sigma \times B(H\rightarrow X)$ in
\pp~collisions at $\sqrt{s}=1.96$~TeV for $m_H=100-200$ GeV/$c^2$.
To facilitate comparisons with the standard model and to accommodate analyses 
with different degrees of sensitivity, we present our results in terms of
the ratio of obtained limits  to  cross section in the SM, as a function of
Higgs boson mass, for test masses for which both experiments have performed 
dedicated searches in different channels.  A value of the combined limit ratio 
which is less than or equal to one indicates that that particular Higgs 
boson mass is excluded at the
95\% C.L.

The combinations of results of each single experiment, as used in this 
Tevatron combination, yield the following ratios of 95\% C.L. observed 
(expected) limits to the SM cross section: 
3.10~(2.38) for CDF and 4.05~(2.80) for D0 at $m_{H}=115$~GeV/$c^2$, and 
1.18~(1.19) for CDF and 1.53~(1.35) for D0 at $m_{H}=165$~GeV/$c^2$.

The ratios of the 95\% C.L. expected and observed limit to the SM
cross section are shown in Figure~\ref{fig:comboRatio} for the
combined CDF and D0 analyses.  The observed and median expected
ratios are listed for the tested Higgs boson masses in
Table~\ref{tab:ratios} for $m_{H} \leq 150$~GeV/$c^2$, and in
Table~\ref{tab:ratios-3} for $m_{H} \geq 155$~GeV/$c^2$, as obtained 
by the Bayesian and the $CL_S$ methods.  In the following summary we 
quote the only limits obtained with the Bayesian method, which was decided
upon {\it a priori}.  It turns out that the Bayesian limits are slightly less
stringent.
The corresponding limits and expected limits obtained using the $CL_S$ 
method are shown alongside the Bayesian limits
in the tables.  We obtain the observed (expected) 
values of 2.70~(1.78) at $m_{H}=115$~GeV/$c^2$, 1.09~(0.96) at 
$m_{H}=160$~GeV/$c^2$, 0.94~(0.89) at $m_{H}=165$~GeV/$c^2$, and 
1.29~(1.07) at $m_{H}=170$~GeV/$c^2$.  
This result is obtained with both Bayesian and $CL_S$ calculations.
\begin{figure}[hb]
\begin{centering}
\includegraphics[width=16.5cm]{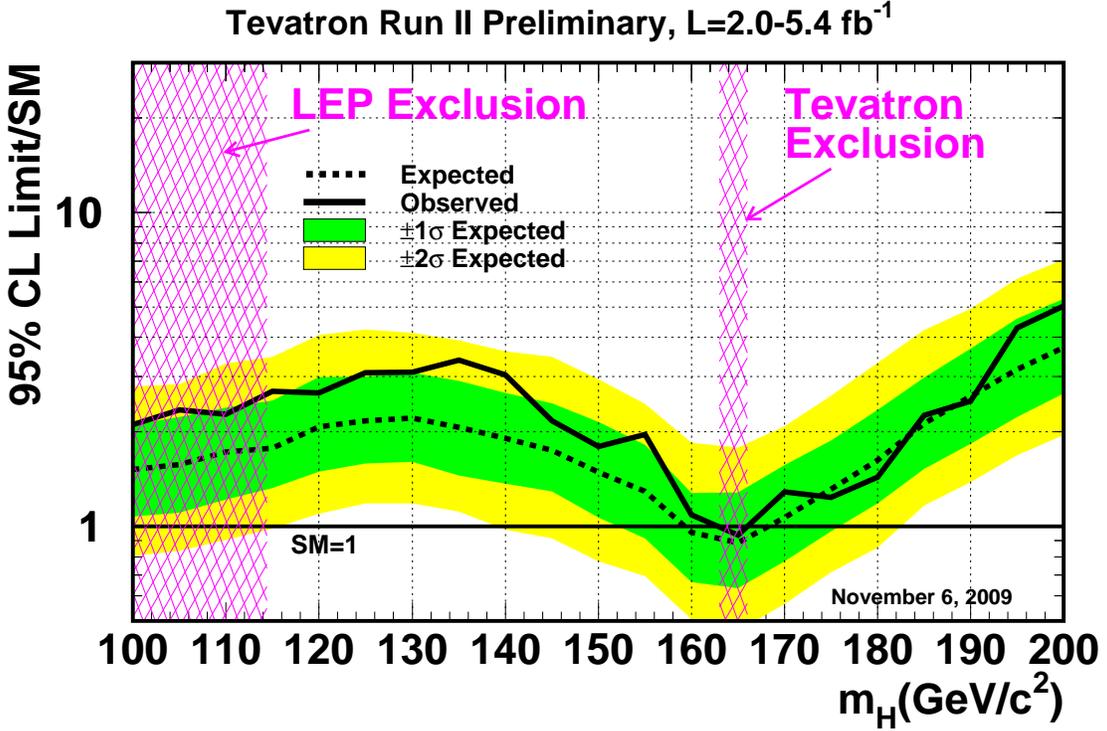}
\caption{
\label{fig:comboRatio}
Observed and expected (median, for the background-only hypothesis)
95\% C.L. upper limits on the ratios to the SM cross section, as 
functions of the Higgs boson mass 
for the combined CDF and D0 analyses.
The limits are expressed as a multiple of the SM prediction
for test masses (every 5 GeV/$c^2$)
for which both experiments have performed dedicated
searches in different channels.
The points are joined by straight lines 
for better readability.
  The bands indicate the
68\% and 95\% probability regions where the limits can
fluctuate, in the absence of signal. 
The limits displayed in this figure
are obtained with the Bayesian calculation.
}
\end{centering}
\end{figure}

\begin{table}[ht]
\caption{\label{tab:ratios} Ratios of median expected and observed 95\% C.L.
limit to the SM cross section for the combined CDF and D0 analyses as a function
of the Higgs boson mass in GeV/$c^2$, obtained with the Bayesian and with the $CL_S$ method.}
\begin{ruledtabular}
\begin{tabular}{lccccccccccc}\\
Bayesian       &  100 &  105 &  110 &  115 &  120 &  125 &  130 &  135 &  140 &  145 &  150 \\ \hline 
Expected       & 1.52 & 1.58 & 1.73 & 1.78 & 2.1 & 2.2 & 2.2 & 2.1 & 1.91 & 1.75 & 1.49 \\
Observed       & 2.11 & 2.35 & 2.28 & 2.70 & 2.7 & 3.1 & 3.1 & 3.4 & 3.03 & 2.17 & 1.80 \\

\hline
\hline\\
$CL_S$         &  100 &  105 &  110 &  115 &  120 &  125 &  130 &  135 &  140 &  145 &  150 \\ \hline 
Expected       & 1.50 & 1.54 & 1.64 & 1.77 & 2.1 & 2.2 & 2.2 & 2.1 & 2.0 & 1.86 & 1.59 \\ 
Observed       & 2.19 & 2.19 & 2.16 & 2.81 & 2.8 & 3.2 & 3.2 & 3.5 & 3.2 & 2.25 & 2.02 \\ 
\end{tabular}
\end{ruledtabular}
\end{table}

\begin{table}[ht]
\caption{\label{tab:ratios-3}
Ratios of median expected and observed 95\% C.L.
limit to the SM cross section for the combined CDF and D0 analyses as a function
of the Higgs boson mass in GeV/$c^2$, obtained with the Bayesian and with the $CL_S$ method.}
\begin{ruledtabular}
\begin{tabular}{lccccccccccc}
Bayesian             &  155 &  160 &  165 &  170 &  175 &  180 &  185 &  190 &  195 &  200 \\ \hline 
Expected             & 1.30 & 0.96 & 0.89 & 1.07 & 1.32 & 1.63 & 2.1 & 2.6 & 3.2 & 3.7 \\
Observed             & 1.96 & 1.09 & 0.94 & 1.29 & 1.24 & 1.44 & 2.3 & 2.5 & 4.3 & 5.0 \\
\hline
\hline\\
$CL_S$               &  155 &  160 &  165 &  170 &  175 &  180 &  185 &  190 &  195 &  200 \\ \hline 
Expected             & 1.27 & 0.92 & 0.89 & 1.09 & 1.32 & 1.64 & 2.1 & 2.6 & 3.2 & 3.7 \\
Observed             & 1.74 & 1.07 & 0.89 & 1.21 & 1.14 & 1.38 & 2.1 & 2.3 & 4.4 & 4.8 \\
\end{tabular}
\end{ruledtabular}
\end{table}

We also show in Figure~\ref{fig:comboLLR-2} and list in Table~\ref{tab:clsVals} the observed 1-$CL_S$ 
and its expected distribution for the background-only hypothesis as a functions of the Higgs boson mass, 
for $m_H\geq 150$~GeV/$c^2$. This is directly interpreted as the level of exclusion of our search.  This 
figure is obtained using the $CL_S$ method. 
We provide 
the Log-likelihood ratio (LLR) values for our combined Higgs boson search, as obtained using the $CL_S$ 
method in Figure~\ref{fig:comboLLR} and Table~\ref{tab:llrVals}.

In summary, we combine all available CDF and D0 results on SM Higgs search,
based on luminosities ranging from 2.1 to 5.4 fb$^{-1}$.
Compared to our previous combination, more data have been added to the existing
channels and analyses have been further optimized
to gain sensitivity. We use the latest parton distribution
functions and $gg \rightarrow H$ theoretical cross sections when
comparing our limits to the SM predictions at high mass.  

The 95\% C.L. upper limits on Higgs boson production are a factor of 2.70~(0.94) 
times the SM cross section for a Higgs boson mass of $m_{H}=$115~(165)~GeV/$c^2$.
Based on simulation, the corresponding median expected upper limits are 1.78~(0.89). 
Standard Model branching ratios, calculated as functions of the Higgs boson mass, 
are assumed.   

We choose to use the intersections of piecewise linear interpolations of our observed and expected rate limits
in order to quote ranges of Higgs boson masses that are excluded and that are expected to be excluded.
The sensitivities of our searches to Higgs bosons are smooth functions of the Higgs boson mass, and depend
most rapidly on the predicted cross sections and the decay branching ratios -- the decay $H\rightarrow W^+W^-$ is the
dominant decay for the region of highest sensitivity.   The mass resolution of the channels is poor due to the
presence of two highly energetic neutrinos in signal events.  We therefore use the linear interpolations to
extend the results from the 5~GeV/$c^2$ mass grid investigated to points in between.  This results in higher expected
and observed interpolated results than if the full dependence of the cross section and branching ratio were included
as well, since the latter produces limit curves that are concave upwards.
The region of Higgs boson masses excluded at the 95\% C.L. thus obtained is $163<m_{H}<166$~GeV/$c^{2}$.  For 
the first time, the Tevatron combined Higgs searches expect to exclude a SM Higgs 
boson for specific masses, assuming no Higgs boson production.
The mass range expected to be excluded is $159<m_{H}<168$~GeV/$c^{2}$.  The excluded region obtained by finding 
the intersections of the linear interpolations of the observed $1-CL_S$ curve shown 
in Figure~\ref{fig:comboLLR-2} is larger than that obtained with the Bayesian calculation.
We choose to quote the exclusion region using the Bayesian calculation.

While new data have been added and the sensitivity of the analyses has thus increased, the observed exclusion region
is slightly smaller than to that obtained in March 2009~\cite{prevhiggs}, which excluded the range
$160<m_{H}<170$~GeV/$c^{2}$.  The new data added have excess signal-like candidates relative to the background predictions
in the high-mass $H\rightarrow W^+W^-$ channels, but the total excess in the new data is 
less than one standard deviation in size.

The results presented in this paper significantly extend the individual limits of each 
collaboration and our previous combination.  The sensitivity of our combined search is sufficient to
exclude a Higgs boson at high mass and is 
expected to grow substantially in the near future with the additional luminosity 
already recorded at the Tevatron and not yet analyzed, and with additional improvements 
of our analysis techniques which will be propagated in the current and future analyses.

 \begin{figure}[t]
 \begin{centering}
 \includegraphics[width=14.0cm]{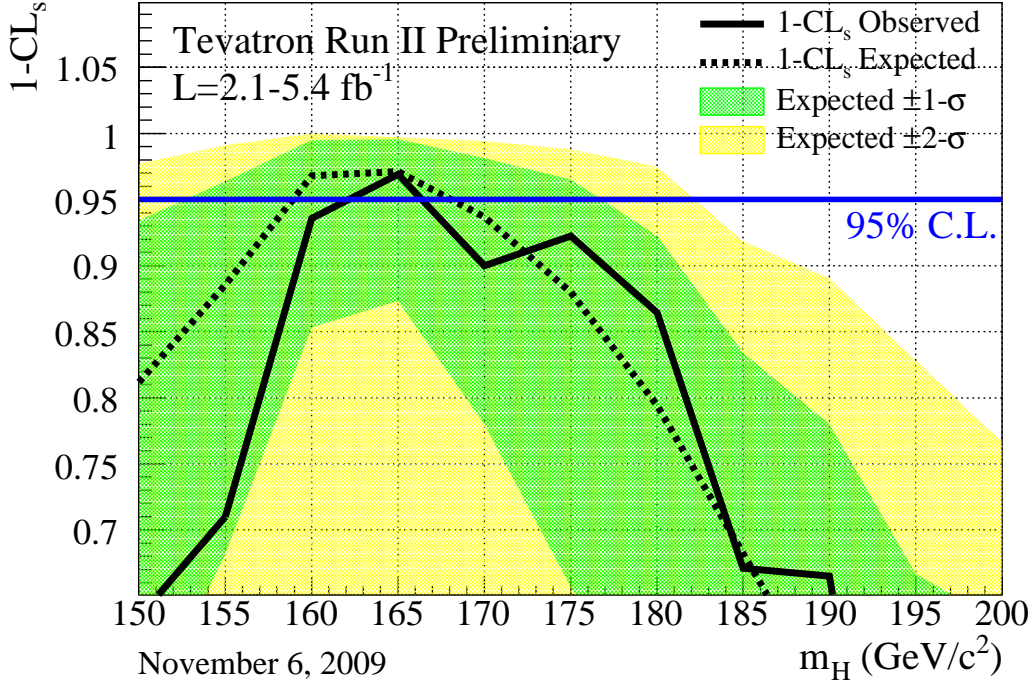}
 \caption{
 \label{fig:comboLLR-2}
 The exclusion strength 1-$CL_S$ as a function of the Higgs boson mass 
(in steps of 5 GeV/$c^2$), as obtained with $CL_S$ method.
 for the combination of the
 CDF and D0 analyses. }
 \end{centering}
 \end{figure}


\begin{table}[htpb]
\caption{\label{tab:clsVals} The observed and expected CL$_{\rm s}$ values as functions of $m_H$, for the combined
CDF and \Dzero Higgs boson searches.}
\begin{ruledtabular}
\begin{tabular}{lcccccc} 
$m_H$ (GeV/$c^2$) & CL$_{\rm s}^{\rm{obs}}$ & 
CL$_{\rm s}^{-2\sigma}$ & 
CL$_{\rm s}^{-1\sigma}$ & 
CL$_{\rm s}^{\rm{median}}$ & 
CL$_{\rm s}^{+1\sigma}$ & 
CL$_{\rm s}^{+2\sigma}$ \\ \hline
100 & 0.562 & 0.968 & 0.931 & 0.829 & 0.571 & 0.258 \\ 
105 & 0.500 & 0.968 & 0.940 & 0.809 & 0.565 & 0.255 \\
110 & 0.563 & 0.948 & 0.914 & 0.765 & 0.506 & 0.212 \\
115 & 0.268 & 0.963 & 0.897 & 0.735 & 0.470 & 0.200 \\
120 & 0.401 & 0.945 & 0.850 & 0.673 & 0.389 & 0.138 \\
125 & 0.300 & 0.894 & 0.815 & 0.646 & 0.391 & 0.137 \\
130 & 0.288 & 0.901 & 0.833 & 0.639 & 0.378 & 0.135 \\
135 & 0.217 & 0.922 & 0.842 & 0.649 & 0.389 & 0.143 \\
140 & 0.283 & 0.915 & 0.862 & 0.693 & 0.412 & 0.158 \\
145 & 0.591 & 0.943 & 0.881 & 0.727 & 0.473 & 0.185 \\
150 & 0.633 & 0.977 & 0.934 & 0.812 & 0.545 & 0.232 \\
155 & 0.710 & 0.991 & 0.964 & 0.886 & 0.682 & 0.370 \\
160 & 0.936 & 1.000 & 0.995 & 0.968 & 0.853 & 0.590 \\
165 & 0.969 & 0.997 & 0.996 & 0.971 & 0.873 & 0.623 \\
170 & 0.900 & 0.994 & 0.981 & 0.937 & 0.780 & 0.478 \\
175 & 0.922 & 0.988 & 0.965 & 0.880 & 0.657 & 0.338 \\
180 & 0.865 & 0.975 & 0.923 & 0.794 & 0.550 & 0.237 \\
185 & 0.671 & 0.918 & 0.834 & 0.682 & 0.407 & 0.145 \\
190 & 0.665 & 0.890 & 0.780 & 0.571 & 0.324 & 0.100 \\
195 & 0.235 & 0.828 & 0.667 & 0.489 & 0.252 & 0.070 \\
200 & 0.272 & 0.766 & 0.629 & 0.439 & 0.219 & 0.059
\end{tabular}
\end{ruledtabular}
\end{table}

\end{document}



\begin{table}[ht]
\caption{\label{tab:ratios} Ratios of median expected and observed 95\% C.L.
limit to the SM cross section for the combined CDF and D0 analyses as a function
of the Higgs boson mass in GeV/$c^2$, obtained with the Bayesian and with the $CL_S$ method.}
\begin{ruledtabular}
\begin{tabular}{lccccccccccc}\\
Bayesian       &  100 &  105 &  110 &  115 &  120 &  125 &  130 &  135 &  140 &  145 &  150 \\ \hline 
Expected       & 1.52 & 1.58 & 1.73 & 1.78 & 2.08 & 2.17 & 2.22 & 2.07 & 1.91 & 1.75 & 1.49 \\
Observed       & 2.11 & 2.35 & 2.28 & 2.70 & 2.66 & 3.09 & 3.10 & 3.39 & 3.03 & 2.17 & 1.80 \\

\hline
\hline\\
$CL_S$         &  100 &  105 &  110 &  115 &  120 &  125 &  130 &  135 &  140 &  145 &  150 \\ \hline 
Expected       & 1.50 & 1.54 & 1.64 & 1.77 & 2.09 & 2.17 & 2.20 & 2.10 & 2.02 & 1.86 & 1.59 \\ 
Observed       & 2.19 & 2.19 & 2.16 & 2.81 & 2.81 & 3.17 & 3.23 & 3.54 & 3.16 & 2.25 & 2.02 \\ 
\end{tabular}
\end{ruledtabular}
\end{table}

\begin{table}
\caption{Systematic uncertainties on the contributions for D0's
$H\rightarrow WW \rightarrow\ell^{\pm}\ell^{\prime \mp}$ channels.
Systematic uncertainties are listed by name; see the original references for a detailed explanation of their meaning and on how they are derived.
Systematic uncertainties shown in this table are obtained for the $m_H=165$ GeV/c$^2$ Higgs selection.
Uncertainties are relative, in percent, and are symmetric unless otherwise indicated.   }
\label{tab:d0systww}
\vskip 0.8cm
{\centerline{D0: $H\rightarrow WW \rightarrow e^{\pm} e^{ \mp}$ Analysis }}
\vskip 0.099cm
\begin{ruledtabular}
\begin{tabular}{ l  c  c  c  c  c  c  c}  \\

Contribution & Diboson & ~~$Z/\gamma^* \rightarrow \ell\ell$~~&$~~W+jet/\gamma$~~ &~~~~$t\bar{t}~~~~$    & ~~Multijet~~  & ~~~~$H$~~~~      \\
\hline
Lepton ID                        &  6           &   6           & 6             & 6            & --   &   6        \\
Charge mis-ID                    &  1           &   1           & 1             & 1            & --   &   1        \\
Jet Energy Scale                 &  1--2     &   1--2        & 1--2          & 1--2          & --   &   1--2       \\
Jet identification               &  1           &   1           & 1             & 1            & --   &   1        \\
Cross Section /normalization      &  7           &   7           & 7            & 10           & 2  &   11         \\
Modeling~~~~~                     &  0           &   1           & 5             & 0            &  --   &   0     \\

\end{tabular}
\end{ruledtabular}
\vskip 0.8cm
{\centerline{D0: $H\rightarrow WW \rightarrow e^{\pm} \mu^{\mp}$ Analysis }}
\vskip 0.099cm
\begin{ruledtabular}
\begin{tabular}{ l  c  c  c  c  c  c  c} \\

Contribution & Diboson & ~~$Z/\gamma^* \rightarrow \ell\ell$~~&$~~W+jet/\gamma$~~ &~~~~$t\bar{t}~~~~$    & ~~Multijet~~  & ~~~~$H$~~~~      \\
\hline
Trigger                          &  2           &   2           & 2             & 2            & --   &   2         \\
Lepton ID                        &  3           &   3           & 3             & 3            & --   &   3        \\
Jet Energy Scale                 &  1           &   1           & 1             & 1            & --   &   1       \\
Jet identification               &  1          &   1           & 1             & 1            & --   &   1        \\
Cross Section/normalization      &  7           &   7           & 7            & 10           & 10   &   11       \\
Modeling~~~~~                    &  0           &   1           & 3             & 0            &  0   &   0      \\

\end{tabular}
\end{ruledtabular}
\label{tab:d0systww_mm}
\vskip 0.8cm
{\centerline{D0: $H\rightarrow WW \rightarrow \mu^{\pm} \mu^{\mp}$ Analysis }}
\vskip 0.099cm
\begin{ruledtabular}
\begin{tabular}{ l  c  c  c  c  c  c  c} \\

Contribution & Diboson & ~~$Z/\gamma^* \rightarrow \ell\ell$~~&$~~W+jet/\gamma$~~ &~~~~$t\bar{t}~~~~$    & ~~Multijet~~  & ~~~~$H$~~~~      \\
\hline
Lepton ID                        &  4           &   4           & 4             & 4            & --   &   4      \\
Momentum resolution              &  1           &   1           & 1             & 1            & --   &   1      \\
Charge mis-ID                    &  1           &   1           & 1             & 1            & --   &   1      \\
Jet identification               &  1           &   1           & 1             & 1            & --   &   1      \\
Cross Section/normalization      &  7           &   7           & 7            & 10           &  15   &   11     \\
Modeling~~~~~                    &  0           &   0           & 1             & 0            &  0   &   0      \\

\end{tabular}
\end{ruledtabular}
\end{table}